\documentclass[final,3p,times,12pt,fixfloat]{elsarticle}

\usepackage[caption=false]{subfig}
\graphicspath{./newplots}

\usepackage{graphicx}
\usepackage{amssymb} 
\usepackage{amsmath} 
\usepackage{times}
\usepackage{color}

\usepackage{txfonts}

\journal{Nuclear Physics B}

\pdfoutput=1

\usepackage{ifpdf}
\ifpdf
\usepackage[pdftex]{hyperref}
\else
\usepackage[hypertex]{hyperref}
\fi

 \hypersetup{
   pdftitle={},%
   pdfauthor={},%
   pdfsubject={},%
   pdfkeywords={},%
   pdfstartview={},%
   bookmarksopen=true, breaklinks=true, debug=true, %
   colorlinks=true, linkcolor=red, citecolor=blue, urlcolor=blue
 }

\usepackage{float} 
\usepackage{subfig} 

\usepackage{amsmath}
\usepackage{epsfig}
\usepackage{float}
\usepackage{verbatim} 
\usepackage{color}

\newcommand{\ie}{{\it i.e.}}
\newcommand{\eg}{{\it e.g.}}

\newcommand{\Q}{{\cal Q}}
\newcommand{\C}{{\cal C}}
\newcommand{\B}{{\cal B}}
\def\soge{{\bigl.^{2s+1}\hspace{-1mm}S^{[8]}_J}}
\def\ss{{\bigl.^3\hspace{-1mm}S^{[1]}_1}}
\def\sps{{\bigl.^1\hspace{-1mm}S^{[8]}_0}}
\def\so{{\bigl.^3\hspace{-1mm}S^{[8]}_1}}
\def\pjs{{\bigl.^3\hspace{-1mm}P^{[1]}_J}}
\def\p0s{{\bigl.^3\hspace{-1mm}P^{[1]}_0}}
\def\p1s{{\bigl.^3\hspace{-1mm}P^{[1]}_1}}
\def\p2s{{\bigl.^3\hspace{-1mm}P^{[1]}_2}}
\def\tpos{{\bigl.^3\hspace{-1mm}P^{[1]}_1}}
\def\tpts{{\bigl.^3\hspace{-1mm}P^{[1]}_2}}
\def\pj{{\bigl.^3\hspace{-1mm}P^{[8]}_J}}
\def\p0{{\bigl.^3\hspace{-1mm}P^{[8]}_0}}

\newcommand{\bq}{\begin{equation}}
\newcommand{\eq}{\end{equation}}
\newcommand{\bqa}{\begin{eqnarray}}
\newcommand{\eqa}{\end{eqnarray}}
\newcommand{\HELACOnia}{{\sc\small HELAC-Onia}}

\def\ss{{\bigl.^3\hspace{-1mm}S^{[1]}_1}}
\def\sps{{\bigl.^1\hspace{-1mm}S^{[8]}_0}}
\def\so{{\bigl.^3\hspace{-1mm}S^{[8]}_1}}
\def\pjs{{\bigl.^3\hspace{-1mm}P^{[1]}_J}}
\def\tpos{{\bigl.^3\hspace{-1mm}P^{[1]}_1}}
\def\tpts{{\bigl.^3\hspace{-1mm}P^{[1]}_2}}
\def\pj{{\bigl.^3\hspace{-1mm}P^{[8]}_J}}
\def\p0{{\bigl.^3\hspace{-1mm}P^{[8]}_0}}

\def\tpzs{{\bigl.^3\hspace{-1mm}P^{[1]}_0}}

\newcommand{\ce}[1]{Eq.~(\ref{#1})}

\newcommand{\cf}[1]{{Fig.~\ref{#1}}}

\newcommand{\ct}[1]{{Tab.~\ref{#1}}}

\usepackage{multirow}

\begin{document}

\hfill CERN-PH-TH-2015-094

\begin{frontmatter}
\title{Double-quarkonium production at a fixed-target experiment at the LHC (AFTER@LHC)}

\author[IPNO]{Jean-Philippe Lansberg}
\author[CERN]{Hua-Sheng Shao}
\address[IPNO]{IPNO, Universit\'e Paris-Sud, CNRS/IN2P3, F-91406, Orsay, France}
\address[CERN]{PH Department, TH Unit, CERN, CH-1211, Geneva 23, Switzerland}

\date{\today}

\begin{abstract}

We present predictions for double-quarkonium production in the kinematical region relevant for the proposed fixed-target experiment using the LHC beams (dubbed as AFTER@LHC). These include all spin-triplet $S$-wave charmonium and bottomonium pairs, i.e. $\psi(n_1S)+\psi(n_2S)$, $\psi(n_1S)+\Upsilon(m_1S)$ and $\Upsilon(m_1S)+\Upsilon(m_2S)$ with $n_1,n_2=1,2$ and $m_1,m_2=1,2,3$. We calculate the contributions from double-parton scatterings and single-parton scatterings. With an integrated luminosity of 20 fb$^{-1}$ to be collected at AFTER@LHC, we find that the yields for double-charmonium production are large enough for differential distribution measurements. We discuss some differential distributions for $J/\psi+J/\psi$ production, which can help to study the physics of double-parton and single-parton scatterings in a new energy range and which might also be sensitive to double intrinsic $c\bar c$ coalescence at large negative Feynman $x$.

\end{abstract}


\end{frontmatter}

\section{Introduction}

Heavy-quarkonium production is typically a multi-scale process, which involves both short- and long-distance facets of the strong interaction.
 This particularity makes heavy-quarkonium production an ideal  probe to study Quantum Chromodynamics (QCD) in its perturbative and 
non-perturbative regimes simultaneously. Studies have extensively been performed at collider and fixed-target energies 
in proton-proton, proton-nucleus and nucleus-nucleus collisions (see reviews e.g. Refs.~\cite{Brambilla:2010cs,Lansberg:2006dh,Andronic:2015wma}). 
The associated production of heavy quarkonium is a very interesting  process not only because it provides a way to pin down the 
heavy-quarkonium production mechanism but also because it can help to understand a new dynamics of hadron collisions appearing 
at high energies, where multiple scatterings of partons
(MPS) happen simultaneously, among which the most likely is of course two short-distance interactions from a single hadron-hadron 
collision -- double-parton scattering (DPS). A number of experimental studies relevant for DPS analyses with heavy quarkonia
have recently been carried out such as $J/\psi+W$~\cite{Aad:2014rua}, $J/\psi+Z$~\cite{Aad:2014kba}, 
$J/\psi+$charm~\cite{Aaij:2012dz} and $J/\psi+J/\psi$~\cite{Abazov:2014qba} production.

In particular, the latter process, \ie\ double-quarkonium production, is of specific interest. It provides an original tool to study the 
quarkonium production from the conventional single-parton scatterings (SPSs), whose contribution has theoretically been studied in many works~\cite{Kartvelishvili:1984ur,Humpert:1983yj,Vogt:1995tf,Li:2009ug,Qiao:2009kg,Ko:2010xy,Berezhnoy:2011xy,Li:2013csa,Lansberg:2013qka,Sun:2014gca,Lansberg:2014swa,Likhoded:2015zna}. 
Moreover, it has been claimed in Refs.~\cite{Kom:2011bd,Baranov:2011ch,Berezhnoy:2012xq,Baranov:2012re,d'Enterria:2013ck,d'Enterria:2014dva,Lansberg:2014swa,Likhoded:2015zna} that DPS contributions should be a significant source of $J/\psi+J/\psi$, 
especially at high energies where there is a high gluon flux. 
On the experimental side, the spin-triplet $S$-waves (\eg\ $J/\psi$, $\psi'$, $\Upsilon(nS)$) provide clean signatures with their small background when they are studied in 
their decay into muon pairs. They are easy to trigger on, in contrast to hadronic jets and open-charm meson productions,
which require either good calorimetry or good  particle identification.

A first comprehensive comparison between experiments~\cite{Aaij:2011yc,Abazov:2014qba,Khachatryan:2014iia} and theory
for $J/\psi$-pair production at the Tevatron and the LHC has been performed in Ref.~\cite{Lansberg:2014swa}, where we 
have pointed out that this observable could be used to probe different mechanisms in different kinematical regions. We noted 
that the direct DPS measurement by D0 collaboration~\cite{Abazov:2014qba} --looking at the rapidity-difference spectrum-- 
is consistent with the $J/\psi$-pair measurement by the CMS collaboration~\cite{Khachatryan:2014iia} and, as we will discuss later on,
compatible with rather large DPS rates. On the other hand, as we advocated in \cite{Lansberg:2013qka},  one cannot draw a definite 
conclusion on the presence of DPS in the early LHCb data~\cite{Aaij:2011yc} with their relatively low statistics.

In this context, we find it important to study the potentialities offered by the use of the 7~TeV proton LHC beams in the fixed-target mode to study quarkonium-pair production. Its multi-TeV beams indeed allow one to study $p+p$, $p+d$ and $p+A$ collisions at 
a centre-of-mass energy $\sqrt{s_{NN}} \simeq 115$~GeV as well as ${\rm Pb}+p$ and ${\rm Pb}+A$ 
collisions at $\sqrt{s_{NN}} \simeq 72$~GeV, with the high precision typical of the 
fixed-target mode.  It has indeed been advocated in~\cite{Brodsky:2012vg,Lansberg:2012kf} that
such a facility, referred to as AFTER@LHC, would become a quarkonium, prompt photon and heavy-flavour 
observatory thanks to its large expected luminosity (for recent phenomenological studies, see~\cite{Liu:2012vn,Boer:2012bt,Chen:2014hqa,Kanazawa:2015fia,Mikkelsen:2015dva,Goncalves:2015hra,Lansberg:2015kha,Ceccopieri:2015rha,Anselmino:2015eoa,Lyonnet:2015dca}). A first feasibility study for quarkonium production was presented in~\cite{Massacrier:2015qba} and
demonstrated that a LHCb-like detector would perform extremely well in the fixed-target mode. Similar performances are expected
for quarkonium-pair production.  

Integrated 
luminosities as large as 20 fb$^{-1}$~\cite{Brodsky:2012vg} can be delivered during a one-year run of $p+{\rm H}$ collisions with 
a bent crystal to extract the beam~\cite{Uggerhoj:2005xz}. The LHC beam 
can also go through an internal-gas-target system\footnote{This is in fact already tested at 
low gas pressures by the LHCb collaboration in order to monitor the 
luminosity of the beam~\cite{Barschel:2014iua,FerroLuzzi:2005em,Aaij:2014ida}.}.
Conservatively sticking to gas pressures already reachable now, yearly integrated  luminosities reach 100 pb$^{-1}$. With a designed 
target cell similar to that of HERMES~\cite{Airapetian:2004yf}, 
a few fb$^{-1}$ yr$^{-1}$ are probably also easily reachable~\cite{Barschel:463141}.  We have reported in \ct{tablumi} the instantaneous and yearly 
integrated luminosities expected with the proton beams on various target species of 
various thicknesses, for both options.

\begin{table}[hbt!]
\begin{center}
{\begin{tabular}{|c c c c c c|}
  \hline
  Beam & Target & Thickness & $\rho$ & $\cal{L}$ & $\int{\cal{L}}$  \\ 
   &  & (cm) & (g.cm$^{-3}$) & ($\mu$b$^{-1}$.s$^{-1}$) & (pb$^{-1}$.y$^{-1})$  \\ \hline
  p & Liquid H & 100 & 0.068 & 2000 & 20000 \\ & & & & &
 \\ \hline \hline
  Beam & Target & Usable gas zone & Pressure &  $\cal{L}$ & $\int{\cal{L}}$    \\
       &  & (cm) & (Bar) & ($\mu$b$^{-1}$.s$^{-1}$) & (pb$^{-1}$.y$^{-1})$  \\ \hline
    p  & perfect gas & 100 & $10^{-9}$ & 10 & 100 \\ \hline
      
  \hline
\end{tabular}}
\caption{\label{tablumi}Expected luminosities obtained for a 7~TeV proton beam extracted by means of a bent crystal or obtained with an internal gas target
with a pressure similar to that of SMOG@LHCb~\cite{FerroLuzzi:2005em}.}
\end{center}
\end{table} 

The structure of this paper is as follows. In section 2, we detail and justify our methodology to compute both DPS and SPS contributions to quarkonium-pair production. Section 3 contains a general discussion of the interest to look at DPS vs SPS contributions at different energies. Section 4 presents a comparison between results up to $\alpha_s^4$ and $\alpha_s^5$. 
This prepares the discussion of our results at $\sqrt{s}=115$ GeV relevant for AFTER@LHC in Section 5. Section 6 gathers our conclusions.

\section{Methodology\label{sec:meth}}

In this section, we explain the main ingredients used to compute the rates for double-quarkonium production at AFTER@LHC, 
which closely follows from our previous work in Ref.~\cite{Lansberg:2014swa}.

\subsection{Double-parton scatterings}

The description of such a mechanism is usually done by assuming that DPSs can be factorised into two single-parton scatterings (SPS)
resulting each in the production of a quarkonium. This can be seen as a first rough approximation which can however be justified by the fact that possible unfactorisable 
corrections due to parton correlations could be small at small $x$. In the case of the double-quarkonium production, 
the master formula from which one starts under the factorisation assumption is~(see e.g. Ref.~\cite{d'Enterria:2013ck})
\bqa
\sigma_{\Q_1\Q_2}&=&\frac{1}{1+\delta_{\Q_1\Q_2}}\sum_{i,j,k,l}{\int{dx_1dx_2dx_1^{\prime}dx_2^{\prime}}d^2{\bold b_1}d^2{\bold b_2}d^2{\bold b}}
\nonumber\\
&&\times \, \Gamma_{ij}(x_1,x_2,{\bold b_1},{\bold b_2}) \, \hat{\sigma}^{\Q_1}_{ik}(x_1,x_1^{\prime})\, \hat{\sigma}^{\Q_2}_{jl}(x_2,x_2^{\prime})\, 
\Gamma_{kl}(x_1^{\prime},x_2^{\prime},{\bold b_1}-{\bold b},{\bold b_2}-{\bold b}) ,
\eqa
where   $\Gamma_{ij}(x_1,x_2,{\bold b_1},{\bold b_2})$ is the generalised double distributions with the longitudinal fractions $x_1$,$x_2$ and the transverse impact parameters ${\bold b_1}$ and ${\bold b_2}$, $\hat{\sigma}^{\Q_i}_{jk}(x_l,x_l^{\prime})$ are the usual partonic cross sections for single quarkonium production and $\delta_{\Q_1\Q_2}$ is the Kronecker delta function. A further factorisation assumption is to decompose $\Gamma_{ij}(x_1,x_2,{\bold b_1},{\bold b_2})$ into a longitudinal part and a transverse part
\bq
\Gamma_{ij}(x_1,x_2,{\bold b_1},{\bold b_2})=D_{ij}(x_1,x_2)T_{ij}({\bold b_1},{\bold b_2}),
\eq
where  $D_{ij}(x_1,x_2)$ is the double-parton distribution functions (dPDF)~\cite{Gaunt:2009re}. Moreover, by ignoring the correlations between partons produced from each hadrons, one can further assume
\bqa
D_{ij}(x_1,x_2)=f_i(x_1)f_j(x_2),\nonumber\\
T_{ij}({\bold b_1},{\bold b_2})=T_i({\bold b_1})T_j({\bold b_2}),
\eqa
where $f_i(x_1)$ and $f_j(x_2)$ are the normal single PDFs. This yields to 
\bqa
\sigma_{\Q_1\Q_2}=\frac{1}{1+\delta_{\Q_1\Q_2}}\sum_{i,j,k,l}{\sigma_{ik\to\Q_1}\sigma_{jl\to\Q_2}}\int{d^2{\bold b}}\!\!
\int{\!T_i({\bold b_1})T_k({\bold b_1}-{\bold b})d^2{\bold b_1}}\!
\int{\!T_j({\bold b_2})T_l({\bold b_2}-{\bold b})d^2{\bold b_2}}.
\eqa
If one also ignores the parton flavour dependence in $T_{i,j,k,l}({\bold b})$ and defines the overlapping function
\bq
F({\bold b})=\int{T({\bold b_i})T({\bold b_i}-{\bold b})d^2{\bold b_i}},
\eq
one reaches the so-called ``pocket formula"
\bq
\sigma_{\Q_1\Q_2}=\frac{1}{1+\delta_{\Q_1\Q_2}}\frac{\sigma_{\Q_1}\sigma_{\Q_2}}{\sigma_{\rm eff}},\label{eq:dpseq}
\eq
where $\sigma_{\Q_1}$ and $\sigma_{\Q_2}$ are the cross sections for respectively single $\Q_1$ and $\Q_2$ production 
and $\sigma_{\rm eff}$ is a parameter to characterise an effective spatial area of the parton-parton interactions via
\bq
\sigma_{\rm eff}=\left[\int{d^2{\bold b}F({\bold b})^2}\right]^{-1}.
\eq
Under these assumptions, it is only related to the initial state and should be independent of the final state. However, the validation of its universality (process independence as well as energy independence) and the factorisation in Eq.(\ref{eq:dpseq}) should be cross checked case by case. In a fact, some factorisation-breaking effects have recently been identified (see \eg\ \cite{Blok:2013bpa,Kasemets:2012pr,Diehl:2014vaa}).
Thanks to its larger luminosity and its probably wide rapidity coverage, AFTER@LHC provides a unique opportunity to probe DPS and to extract $\sigma_{\rm eff}$ from double-quarkonium final states.

To perform our predictions, we will use $\sigma_{\rm eff}=5.0\pm2.75$ mb, which was determined from $J/\psi$-pair production data at the Tevatron by D0 collaboration~\cite{Abazov:2014qba}.\footnote{Note that Ref.~\cite{Abazov:2014qba} has updated the value of $\sigma_{\rm eff}$ to be $4.8\pm2.55$ mb. However, since the difference is very small, we still used the original one.} The reason for such a choice is that all of the double-quarkonium-production processes share the same gluon-gluon initial states and the typical $x$ are not that much different. This also means that we only need to assume the energy independent of $\sigma_{\rm eff}$. However, we do not claim that this value is the only one possible; we  only take it as our reference number. If one wants to use another value of $\sigma_{\rm eff}$, one can just simply perform a rescaling (proportional to $1/\sigma_{\rm eff}$) of the numbers given in the following.

\begin{table}[!hbtp] 
\begin{center}
\subfloat[Charmonia]{
\begin{tabular}{c|cccc}\footnotesize
          & $\kappa$ &$\lambda$ & \# of data & $\chi^2$ \\
\hline\hline
$J/\psi$ & $0.67\pm0.08$  & $0.38$ & $51$ & $422$\\
$\psi(2S)$  &  $0.15\pm0.03$ & $0.35$ & $4$ & $1.12$ \\
\\
\end{tabular}\label{cfit}
}       \\
\subfloat[Bottomonia]{
\begin{tabular}{c|cccc}\footnotesize
          & $\kappa$ &$\lambda$ & \# of data & $\chi^2$ \\
\hline\hline
$\Upsilon(1S)$ & $0.89$  & $0.084\pm0.0061$ & $14$ & $29$\\
$\Upsilon(2S)$  &  $0.79$ & $0.056$ & $9$ & $2.2$ \\
$\Upsilon(3S)$  &  $0.68\pm 0.029$ & $0.046$ & $9$ & $3.9$ \\
\end{tabular}\label{bfit}
}
\caption{Results of a fit of $d^2\sigma/dP_Tdy$ to (a) the $\psi(nS)$ 
PHENIX data~\cite{Adare:2011vq} by fixing $n=2$ and $\langle P_T\rangle = 4.5$~GeV and (b) the $\Upsilon(nS)$ data CDF~\cite{Acosta:2001gv} data by fixing $n=2$ and $\langle P_T\rangle = 13.5$ GeV. Only the $>1\%$ errors are given.}
\end{center}
\end{table}

Since the description of single heavy-quarkonium production at hadron colliders in the whole kinematical region is still a challenge to theorists,  using {\it ab initio} theoretical computation of $\sigma_{\Q}$ would significantly inflate theoretical uncertainties. Instead, we will work in a data-driven way to determine $\sigma_{\Q}$.

\begin{figure}[hbt!]
\begin{center}
\subfloat[]{\includegraphics[width=0.49\textwidth,draft=false]{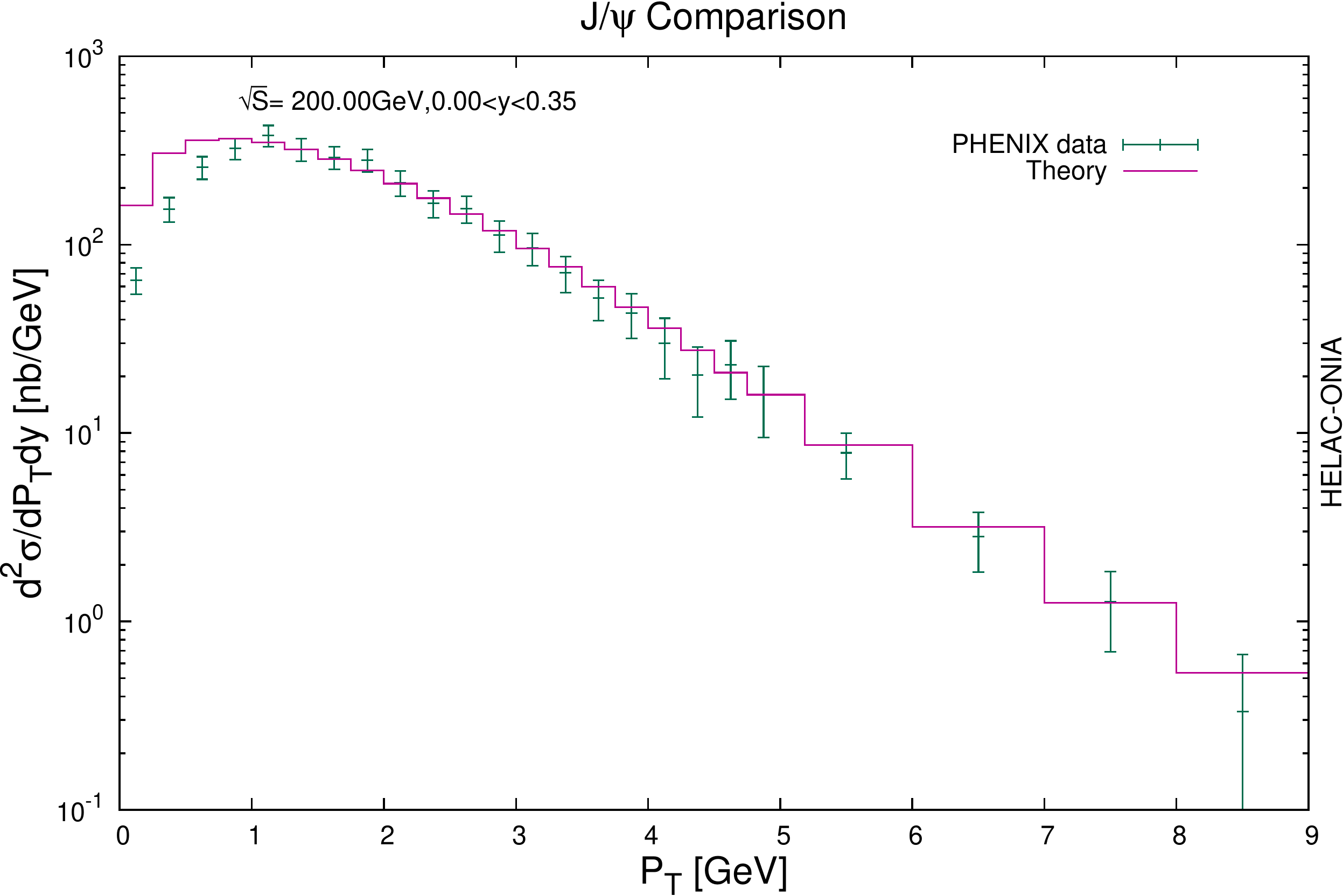}\label{fig:CompareDataJPsi1}}\
\subfloat[]{\includegraphics[width=0.49\textwidth,draft=false]{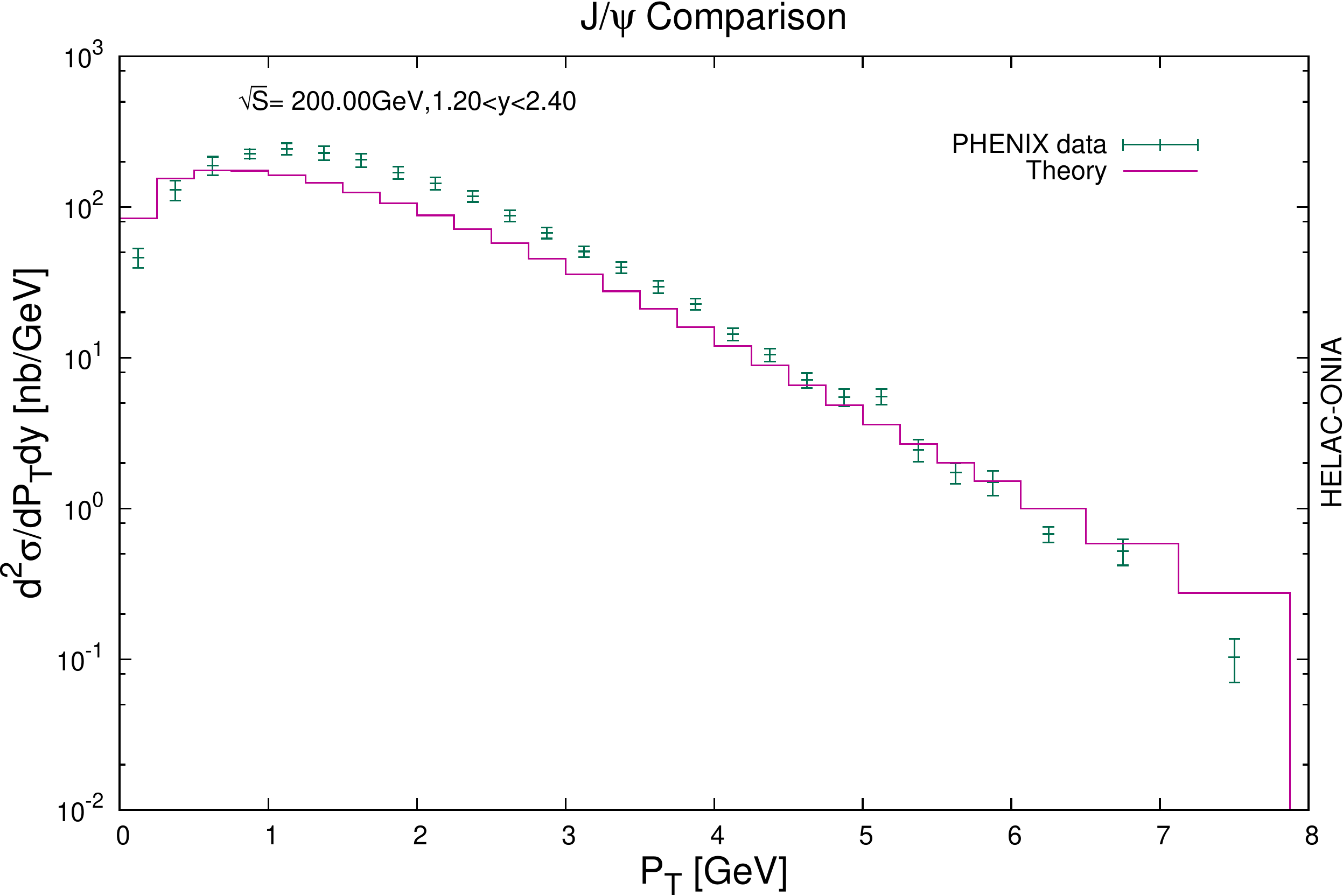}\label{fig:CompareDataJPsi2}}\\
\subfloat[]{\includegraphics[width=0.49\textwidth,draft=false]{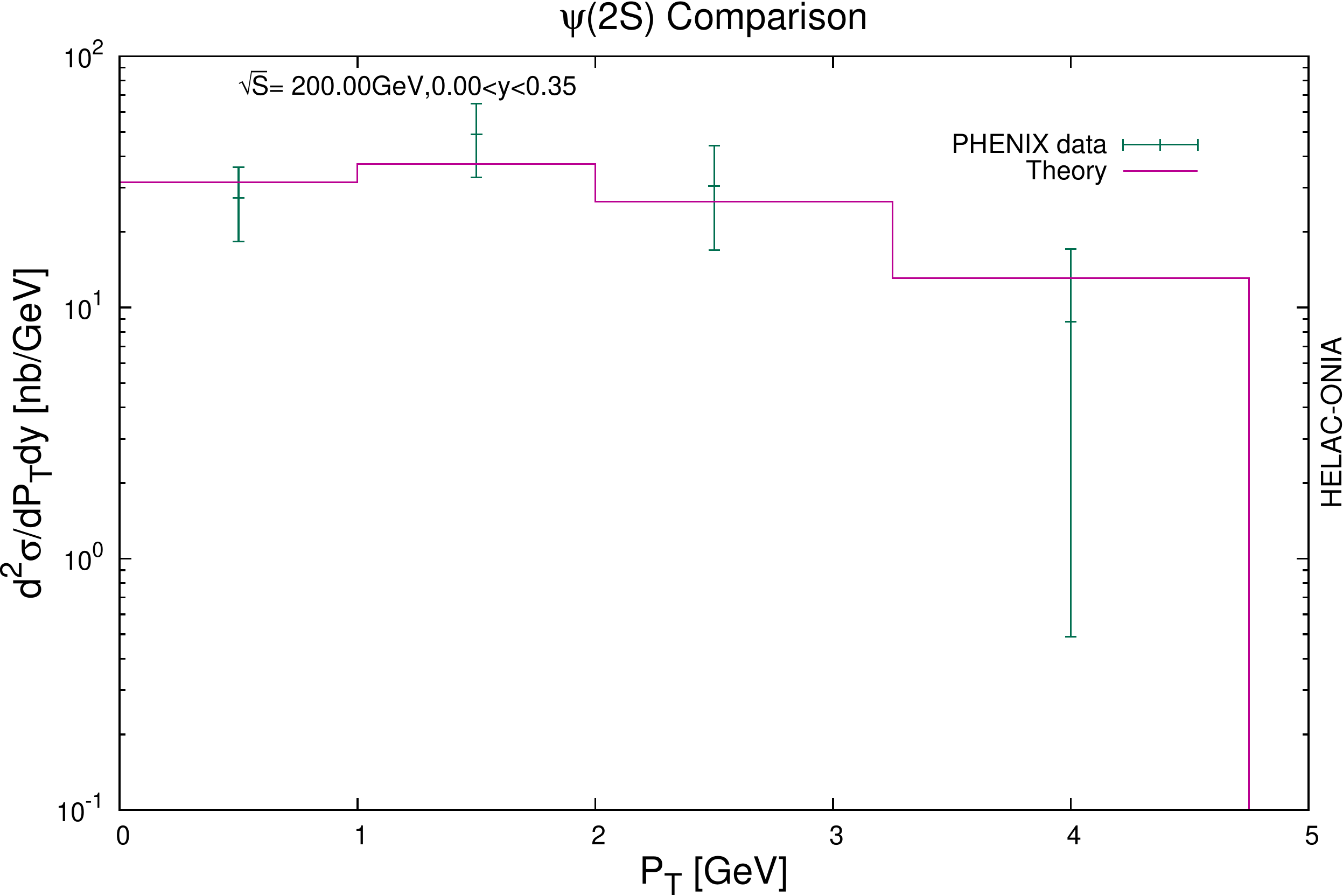}\label{fig:CompareDataPsi2S}}\
\subfloat[]{\includegraphics[width=0.49\textwidth,draft=false]{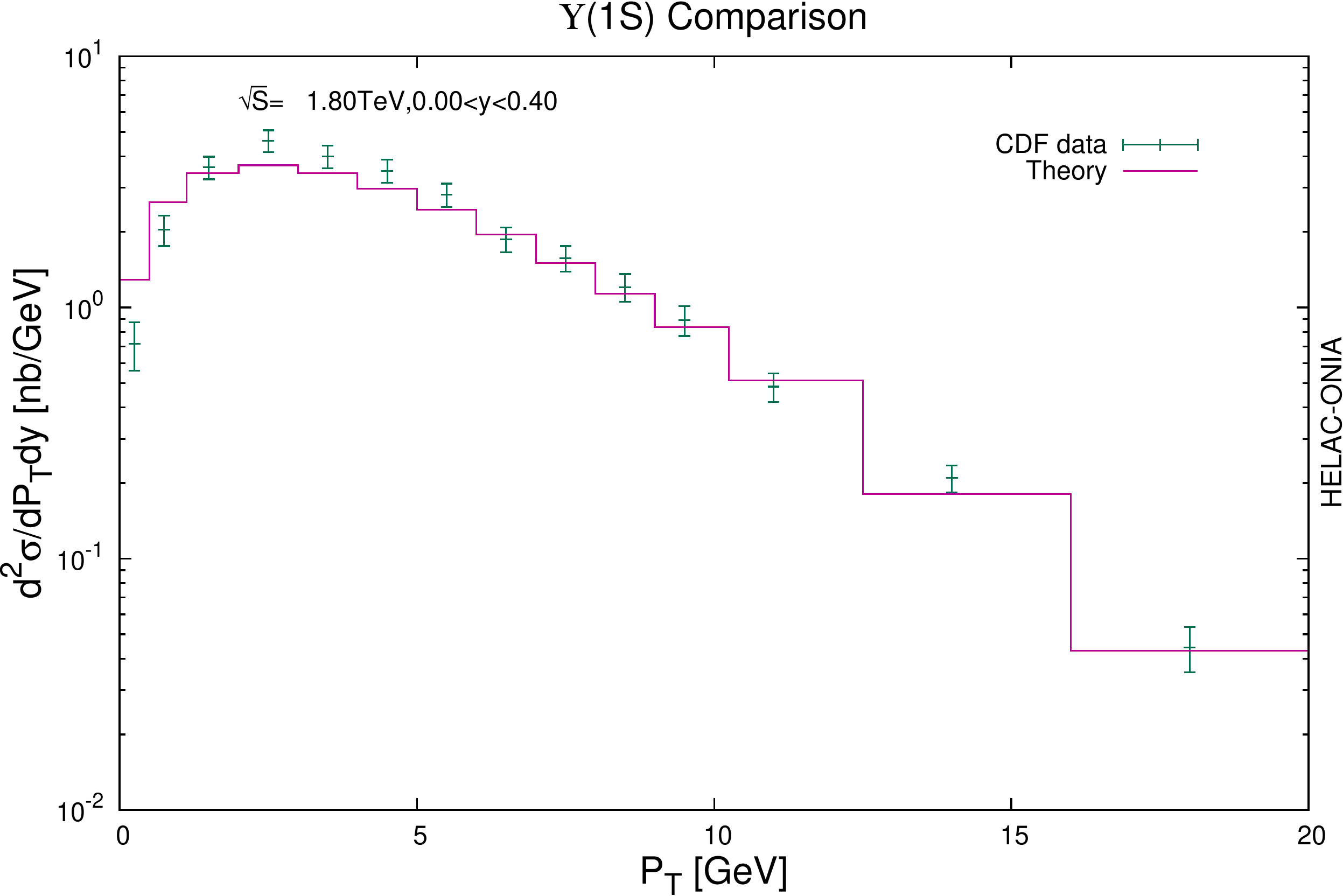}\label{fig:CompareDataUpsi1}}\\
\subfloat[]{\includegraphics[width=0.49\textwidth,draft=false]{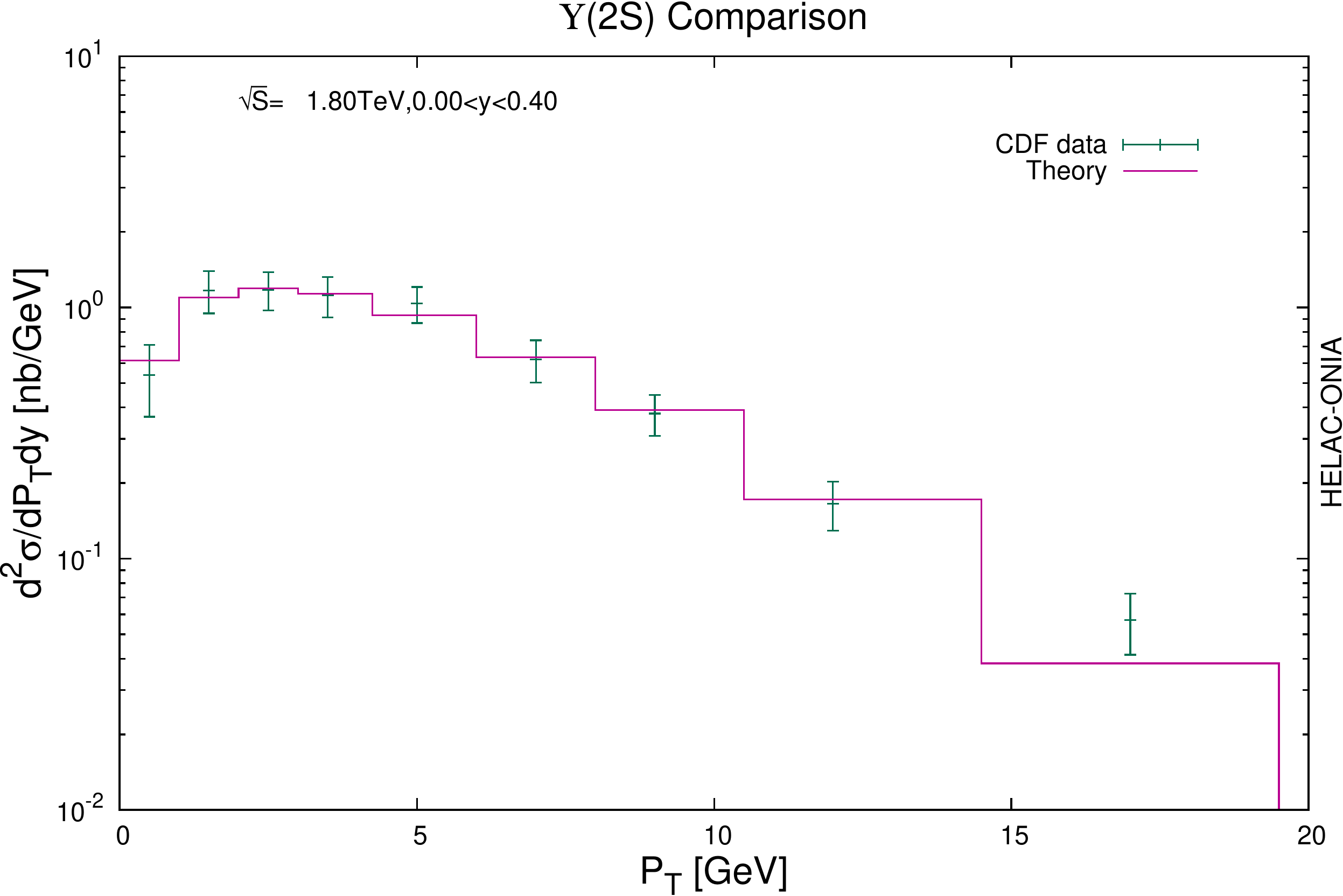}\label{fig:CompareDataUpsi2}}\
\subfloat[]{\includegraphics[width=0.49\textwidth,draft=false]{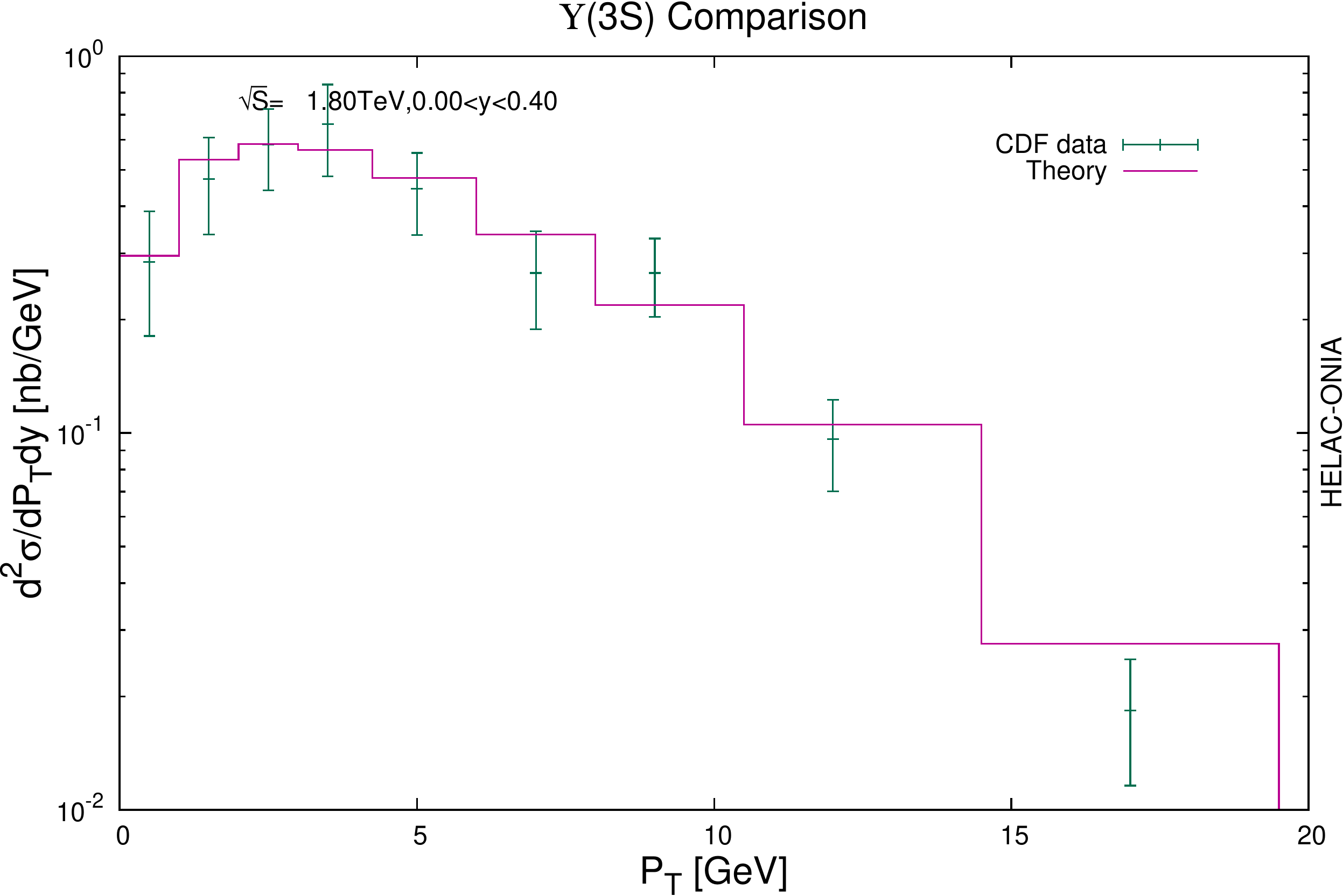}\label{fig:CompareDataUpsi3}}
\caption{Comparisons with the PHENIX measurements~\cite{Adare:2011vq} for $J/\psi$ (a,b) and $\psi(2S)$ (c) production and with the CDF measurements~\cite{Acosta:2001gv} for $\Upsilon(1S)$ (d), $\Upsilon(2S)$ (e) and $\Upsilon(3S)$ (f) production.
}
\label{fig:CompareData}
\end{center}\vspace*{-1cm}
\end{figure}

Our procedure is as follows. We start from the cross section $\sigma_{\Q_i}$ which can be written as
\bqa
\sigma(pp\to\Q+X)&=&\sum_{a,b}\int{dx_1dx_2f_a(x_1)f_b(x_2)}
\frac{1}{2\hat{s}}\overline{|\mathcal{A}_{ab\to\Q+X}|^2}d{\rm LIPS}_{\Q+X},
\eqa
where $f_a,f_b$ are the parton distribution functions (PDF) of the initial partons $a$ and $b$, $d{\rm LIPS}_{\Q+X}$ is the Lorentz-invariant phase-space measure for $pp\to \Q+X$ and $\sqrt{\hat{s}}$ is the partonic centre-of-mass energy (i.e. $\hat{s}=x_1x_2s$). For single quarkonium production in $p+p$ collisions at  $\sqrt{s}=115$ GeV, the gluon-gluon initial state is dominant. The initial colour and helicity averaged amplitude square for $gg\to \Q+X$ can be expressed in the form of a crystal ball function~\cite{Kom:2011bd}
\bqa
&&\overline{|\mathcal{A}_{gg\to\Q+X}|^2}=
\left\{
\begin{array}{ll}
K\exp(-\kappa\frac{P_T^2}{M_{\Q}^2})
& \mbox{when $P_T\leq \langle P_T\rangle$} \\
K\exp(-\kappa\frac{\langle P_T \rangle^2}{M_{\Q}^2})\left(1+\frac{\kappa}{n}\frac{P_T^2-\langle P_T \rangle^2}{M_{\Q}^2}\right)^{-n}
& \mbox{when $P_T> \langle P_T\rangle$} \\
\end{array}
\right.\label{eq:crystalball}
\eqa
where $K=\lambda^2\kappa\hat{s}/M_{\Q}^2$. The parameters $\kappa$,$\lambda$,$n$ and $\langle P_T \rangle$ can be determined by fitting the (differential) cross sections to the experimental data. The dedicated codes to perform the fit and to compute the DPS contributions to double-quarkonium production have been implemented in \HELACOnia~\cite{Shao:2012iz,Shao:2015vga}. 

Once a fit is done, $|\mathcal{A}_{gg\to\Q+X}|^2$ is fixed and it allows us to evaluate $\sigma(pp\to\Q+X)$ (or its differential 
counterparts in any variable) which can then be injected into the ``pocket formula'' \ce{eq:dpseq} in order to predict the 
DPS yield. Since we do not apply any muon cuts, we do not need to make any assumptions regarding the polarisation of 
the production quarkonia.

The code was tested and, with the same parameters as in Ref.~\cite{Kom:2011bd}, we have reproduced their results. 
However, their combined fit of the charmonium data taken at the Tevatron and the LHC cannot  reproduce well 
the low-energy data measured by PHENIX collaboration~\cite{Adare:2011vq} at RHIC. Since the collision energy of 
RHIC $\sqrt{s}=200$ GeV is very close to the centre-of-mass energy of the fixed-target experiment at the LHC 
(AFTER@LHC), \ie\ $\sqrt{s}=115$ GeV, we prefer to use the PHENIX data alone to determine the parameters 
in~\ce{eq:crystalball}. A fit of $d^2\sigma/dP_Tdy$ to the PHENIX data~\cite{Adare:2011vq} for $J/\psi$ and 
$\psi(2S)$ production gives the $\chi^2$ results presented in \ct{cfit} having fixed  $n=2$ and 
$\langle P_T\rangle = 4.5$ GeV. We also show the comparisons of the $P_T$ spectra in \cf{fig:CompareData}a-c. 
The large $\chi^2$ for the single $J/\psi$ production can be reduced to $55.8$ when one only considers the $23$ 
PHENIX data points in the central region (i.e. $|y_{J/\psi}|<0.35$) and excluding the lowest-$P_T$ bin. A fit to the 
sole PHENIX data in the forward region $1.2<|y_{J/\psi}|<2.4$ changes $\kappa$ by $\sim 15\%$ and $\lambda$ 
by $\sim 5\%$. However, the main uncertainty in predicting DPS contributions to  double $\psi$ production 
remains from that of $\sigma_{\rm eff}$ and those from these fits are in practice nearly irrelevant for our 
predictions. This is obvious for $\lambda$ which only affects the normalisation.

In contrast, there is no differential measurement of $\Upsilon$ yields at RHIC. There exists 
data from the fixed-target Fermilab experiment E866~\cite{Zhu:2007aa} but only at low $P_T$. 
We therefore performed a fit of $d^2\sigma/dP_Tdy$ to CDF~\cite{Acosta:2001gv} Run I 
data at $\sqrt{s}=1.8$ TeV.  The results for $\Upsilon$ are presented in \ct{bfit} having fixed $n=2$ 
and $\langle P_T\rangle = 13.5$ GeV. For illustration, the comparisons between the fit and the CDF 
data~\cite{Acosta:2001gv} are shown in \cf{fig:CompareData}d-f. Some comments about the fit are however in order. 
If we instead performed a combined fit to CDF~\cite{Acosta:2001gv}, ATLAS~\cite{Aad:2012dlq}, 
CMS~\cite{Chatrchyan:2013yna} and LHCb~\cite{LHCb:2012aa,Aaij:2013yaa} data, the value of 
$\kappa$ ($\lambda$) would be shifted by at most $30\%$ ($10\%$) but with significantly worse $\chi^2$. 
All this may however not be so relevant since, as for the charmonia, the fit to TeV data tend to underestimate
 the RHIC $P_T$-integrated $\Upsilon$ production cross section as measured by STAR~\cite{Adamczyk:2013poh}
by a factor a bit smaller than 2 -- the STAR result has however a 30\% uncertainty.
The uncertainties on $\kappa$ and $\lambda$ given by the $\chi^2$ fit are therefore far too optimistic
since the Crystall Ball parametrisation seems not to correctly capture  the energy dependence of the cross section.
The corresponding DPS yields of $\Upsilon$ at AFTER@LHC which we give here should therefore be considered as conservative
{\it lower} estimates.  
All of the above fits are performed with MSTW2008NLO PDF set~\cite{Martin:2009iq} available in LHAPDF5~\cite{Whalley:2005nh} 
and the factorisation scale $\mu_F=\sqrt{M_{\Q}^2+P_T^2}$. The physical mass $M_{\Q}$ for quarkonium  is 
taken from PDG data~\cite{Agashe:2014kda} as well as the branching ratios.

\subsection{Single-parton scatterings}

\subsubsection{Double-charmonium and double-bottomonium production}

The SPS contribution to $J/\psi$-pair production have  systematically been investigated in our 
previous works~\cite{Lansberg:2013qka,Lansberg:2014swa}. We have shown that a leading order (LO) calculation in the 
strong coupling constant, $\alpha_s$,  is enough to account for the low-$P_T$ data as well as the 
$P_T$-integrated cross section, the bulk of the events lying at low $P_T$. 
However, if one goes to mid $P_T$ (\eg\ $P_T>5$ GeV), $\mathcal{O}(\alpha_s^5)$ contribution start to be large. 
As a consequence, the yield and the polarisation  changes significantly compared to a LO calculation. 
Since we are only interested in the data which are measurable with up to 20 fb$^{-1}$ in order to assess the feasibility of 
measuring quarkonium-pair production with AFTER@LHC, we will focus on  the low $P_T$ region. As we will explicitly show, 
LO evaluations happen to be sufficient. Besides, the colour-octet contributions are also negligible 
at low $P_T$ for they are suppressed by powers of $v$ without any 
kinematical enhancement at variance with the single-quarkonium-production case. 

\begin{table}[!hbtp] 
\begin{center}
\subfloat[Decay within a family]{
\begin{tabular}{{c}*{1}{c}}\hline\hline
decay channel & branching ratio ($\%$)\\\hline 
$\psi(2S)\to J/\psi+X$ & $57.4$\\
$\Upsilon(2S)\to \Upsilon(1S)+X$ & $30.2$\\
$\Upsilon(3S)\to \Upsilon(1S)+X$ & $8.92$\\
$\Upsilon(3S)\to \Upsilon(2S)+X$ & $10.6$\\\\
\hline\hline
\end{tabular}} \quad
\subfloat[Leptonic decays]{
\begin{tabular}{{c}*{1}{c}}\hline\hline
decay channel & branching ratio ($\%$)\\\hline 
$J/\psi\to \mu^+\mu^-$ & $5.93$\\
$\psi(2S)\to\mu^+\mu^-$ & $0.75$\\
$\Upsilon(1S)\to \mu^+\mu^-$ & $2.48$ \\
$\Upsilon(2S)\to \mu^+\mu^-$ & $1.93$ \\
$\Upsilon(3S)\to \mu^+\mu^-$ & $2.18$ \\
\hline\hline
\end{tabular}}
\end{center}
\caption{Various decays (and branching ratios) considered in this article~\cite{Agashe:2014kda}.}
\label{tab:br}
\end{table}

On the contrary, the feed-down contributions from 
higher excited spin-triplet $S$-wave quarkonium has to be considered. It is  substantial as already shown for the $J/\psi$-pair production in Ref.~\cite{Lansberg:2014swa}. These will systematically be taken into account in our predictions as done
in Ref.~\cite{Lansberg:2014swa}.
The branching ratios that will be used in this context are taken from PDG~\cite{Agashe:2014kda} and we have listed them in 
\ct{tab:br} for completeness.

The general formula for the amplitude of the production of a pair of colour-singlet (CS) $S$-wave quarkonia $\Q_1$ and $Q_2$ with as initial partons $a$ and $b$ is
\bqa
&&\mathcal{A}_{ab\to \Q_1^{\lambda_1}(P_1)+\Q_2^{\lambda_2}(P_2)+X}=\\
&&\sum_{s_1,s_2,c_1,c_2}{\sum_{s_3,s_4,c_3,c_4}{\frac{N(\lambda_1|s_1,s_2)N(\lambda_2|s_3,s_4)}{\sqrt{M_{\Q_1}M_{\Q_2}}}\frac{\delta_{c_1c_2}\delta_{c_3c_4}}{N_c}\frac{R_1(0)R_2(0)}{4\pi}}}\mathcal{A}_{ab\to Q_{c_1}^{s_1}\bar{Q}_{c_2}^{s_2}({\bold p_1}={\bold 0})+Q_{c_3}^{s_3}\bar{Q}_{c_4}^{s_4}({\bold p_2}={\bold 0})+X},\nonumber\label{eq:form}
\eqa
where we denote the momenta of quarkonia $\Q_1$ and $\Q_2$ as $P_1$ and $P_2$ respectively and their polarisations as $\lambda_{1,2}$, $N(\lambda_{1,2}|s_{1,3},s_{2,4})$ are the two spin projectors and $R_{1,2}(0)$ are the radial wave functions at the origin in the configuration space for both quarkonia. In the above equation, we have defined the heavy-quark momenta to be $q_{1,2,3,4}$ such that $P_{1,2}=q_{1,3}+q_{2,4}$ and $p_{1,2}=(q_{1,3}-q_{2,4})/2$. $s_{1,2,3,4}$ are then the heavy-quark spin components and $\delta_{c_ic_j}/\sqrt{N_c}$ is the colour projector. The spin-triplet projector $N(\lambda|s_i,s_j)$ has, in the non-relativistic limit, $v \to 0$, the following expression 
\bq
N(\lambda|s_i,s_j)=\frac{\varepsilon^{\lambda}_{\mu}}{2\sqrt{2}M_{\Q}}\bar{v}(\frac{{\bold P}}{2},s_j)\gamma^{\mu}u(\frac{{\bold P}}{2},s_i).
\eq
All these computations can be performed automatically in the \HELACOnia~\cite{Shao:2012iz} framework based on recursion relations. The radial wave functions at the origin $R(0)$ are taken from Ref.~\cite{Eichten:1995ch}, which were derived in the QCD-motivated Buchm\"uller-Tye potential~\cite{Buchmuller:1980su}. We also listed their values in \ct{tab:R02}.

\begin{table*}[!hbtp] 
\begin{center}
\begin{tabular}{{c}*{1}{c}}\hline\hline
Quarkonium & $|R(0)|^2$ (GeV$^3$)\\\hline 
$J/\psi$ & $0.81$ \\
$\psi(2S)$ & $0.529$ \\
$\Upsilon(1S)$ & $6.477$\\
$\Upsilon(2S)$ & $3.234$\\
$\Upsilon(3S)$ & $2.474$\\
\hline\hline
\end{tabular}
\end{center}
\caption{The radial wave functions at the origin squared $|R(0)|^2$~\cite{Eichten:1995ch} of $S$-wave quarkonium used in this article.}
\label{tab:R02}
\end{table*}

\subsubsection{Charmonium-bottomonium pair production}

The simultaneous production of a charmonium  and a bottomonium has been studied in Refs.~\cite{Ko:2010xy,Likhoded:2015zna}. Its CSM contributions are expected to be suppressed because the direct LO contributions in CS mechanism (CSM) are $\mathcal{O}(\alpha_s^6)$, i.e. $\alpha_s^2$ suppressed compared to double-charmonium and double-bottomonium production. Hence, it is expected to be a golden channel to probe colour-octet mechanism (COM) at the LHC~\cite{Ko:2010xy}. However, such a statement is valid only if one can clearly separate DPS and SPS events experimentally since the DPS contributions would be substantial. For a thorough discussion, the reader is guided to~\cite{Likhoded:2015zna}. In contrast, colour octet (CO) contributions can appear at $\mathcal{O}(\alpha_s^4)$, which however are suppressed by the small size of the CO long distance matrix elements (LDMEs). If one follows the arguments of Ref.~\cite{Ko:2010xy}, one is entitled to consider only the  $c\bar{c}(\so)+b\bar{b}(\so)$, $c\bar{c}(\ss)+b\bar{b}(\so)$ and $c\bar{c}(\so)+b\bar{b}(\ss)$ channels. This approximation is however based on the validity of the velocity
scaling rules of the LMDEs which may not be reliable. A complete computation --even at LHC energies--  
accounting for all the possible channels up to $v^7$ in NRQCD is still lacking in the literature: there are indeed 
more than 50 channels at LO in $\alpha_s$  contributing to $\psi+\Upsilon$ production. Thanks to the the automation 
of \HELACOnia~\cite{Shao:2012iz,Shao:2015vga}, such a complete is at reach.

The formula for the $S$-wave CO amplitude is similar to that for CS state production with the following formal 
replacements for CO in Eq.(\ref{eq:form})
\bq
\frac{\delta_{c_i,c_j}}{\sqrt{N_c}}\to \sqrt{2}T^a_{c_ic_j},
\frac{R_i(0)}{\sqrt{4\pi}}\to  \frac{\sqrt{\langle\mathcal{O}^i(\soge) \rangle}}{\sqrt{(2J+1)(N_c^2-1)}},
\eq
where $T^a_{c_ic_j}$ is the Gell-Mann matrix and $\langle\mathcal{O}^i(\so) \rangle$ is the CO LDME. We refer 
the reader to Ref.~\cite{Shao:2012iz} for the $P$-wave amplitudes.

The non-perturbative CO LDMEs should be determined from experimental data. Their values unfortunately depend much 
on the fit procedures. We took four sets of LDMEs from the literature (see the details in \ref{appA2}).

Finally, we describe our parameters for our SPS calculations. In the non-relativistic limit, the mass of the heavy quarkonium can be expressed as the sum of the corresponding heavy-quark-pair masses. In our case, we have
\bq
M_{\Q}=2m_Q,
\eq
where $m_Q=m_c$ for charmonium and $m_Q=m_b$ for bottomonium. The masses of charm quark and bottom quark are taken as $m_c=1.5\pm0.1$ GeV and $m_b=4.75\pm 0.25$ GeV. The factorisation scale $\mu_F$ and the renormalisation scale $\mu_R$ are taken as $\mu_F=\mu_R \in [\frac{1}{2}\mu_0,2\mu_0]$ with $\mu_0=\sqrt{(M_{\Q_1}+M_{\Q_2})^2+P_T^2}$. The advantage of using $\mu_0=\sqrt{(M_{\Q_1}+M_{\Q_2})^2+P_T^2}$ is that we are able to recover the correct mass threshold $M_{\Q_1}+M_{\Q_2}$ in the low $P_T$ regime. Finally, the PDF set for the SPS calculation is CTEQ6L1~\cite{Pumplin:2002vw} with the one-loop renormalisation group running of $\alpha_s$.

\section{Energy dependence of the ratio DPS over SPS}

Due to the very large integrated luminosity of AFTER@LHC (up to 20 fb$^{-1}$ per year) compared to the experiments performed at RHIC, the measurement of double-quarkonium production at AFTER@LHC will provide a unique test of the interplay between the DPS and SPS production mechanisms in a new energy range. The energy dependence of $\sigma_{\rm eff}$ will be explored at a wide energy range when combined with the LHC collider and Tevatron data\footnote{Since we noted that the energy dependence obtained with the partonic amplitude ($gg \to \Q X)$ given by a Crystal Ball fit with fixed parameters is not optimal when going to TeV energies down to RHIC energies, we have used the fit parameters of~\cite{Kom:2011bd} (based on a fit of Tevatron and LHC data) to predict the DPS yield in the TeV range and our fit to the PHENIX data for the RHIC and fixed-target-experiment energy range.}. Due to the double enhancement of the initial gluon-gluon luminosity with the energy, $\sqrt{s}$, DPS contributions are expected to be more and more important with respect to the SPS ones at larger $\sqrt{s}$. This can be observed on \cf{fig:energydep}.

One however sees on \cf{fig:energydep} that a change of $\sigma_{\rm eff}$ from 15 mb --which seems to be the favoured value for 
jet-related observables-- to 5mb --which is the value extracted by D0 from the $J/\psi+J/\psi$ data~\cite{Abazov:2014qba}-- results in a 
significant change in the point where both contributions are equal. In the former case, it occurs very close to the energy of AFTER@LHC, 
in the latter case, it occurs between the Tevatron and the LHC energies. All this clearly motivates for measurement and $\sigma_{\rm eff}$  
extractions at low energies. 

\begin{figure*}[t!]
\begin{center}
\includegraphics[width=0.7\textwidth,draft=false]{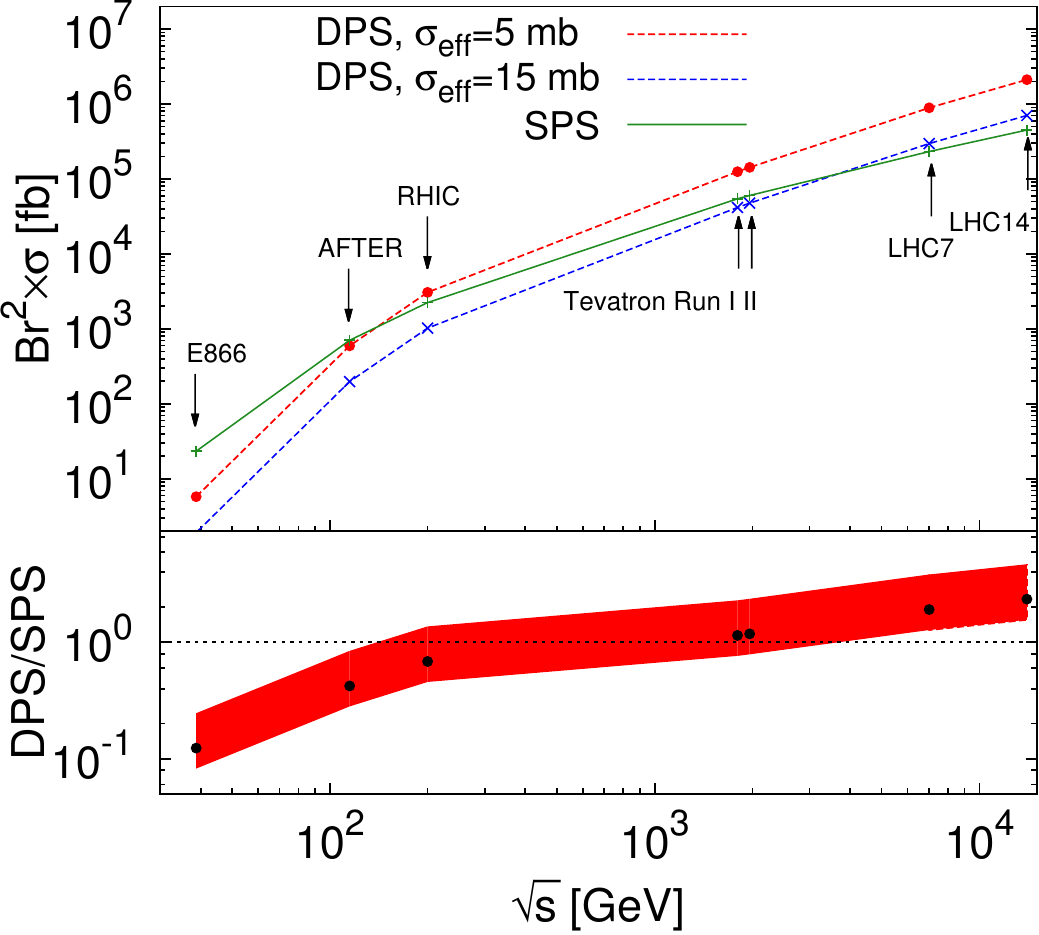}
\caption{(Upper panel) The cross sections of (prompt-)$J/\psi$ pair production via SPS and DPS mechanisms for two values of $\sigma_{\rm eff}$ as a function of $\sqrt{s}$. (Lower panel)
DPS over SPS yield ratio for $5 < \sigma_{\rm eff} < 15$~mb. The black circles correspond to 10 mb. [Aside from the choice of $\sigma_{\rm eff}$, no theoretical uncertainties are included].
}
\label{fig:energydep}
\end{center}\vspace*{0cm}
\end{figure*}

\section{Impact of the QCD corrections at low transverse momenta}

Before showing our results and in order to motivate the use of LO predictions for this exploratory study, we have found it useful to give an explicit
comparison between the differential cross section at LO and NLO$^\star$ for double-$J/\psi$ production 
in the kinematical domain accessible with 20 fb$^{-1}$, that is up to transverse momenta on the order of 10 GeV at the very most.
Indeed, in a previous study~\cite{Lansberg:2013qka}, we have showed that the impact of the real-emission corrections, such as $gg\to J/\psi + J/\psi + g$, becomes increasingly important at large transverse momenta.

\begin{figure}[t!]
\begin{center}
\subfloat[Absolute rapidity difference between both $J/\psi$]{\includegraphics[width=0.49\textwidth,draft=false]{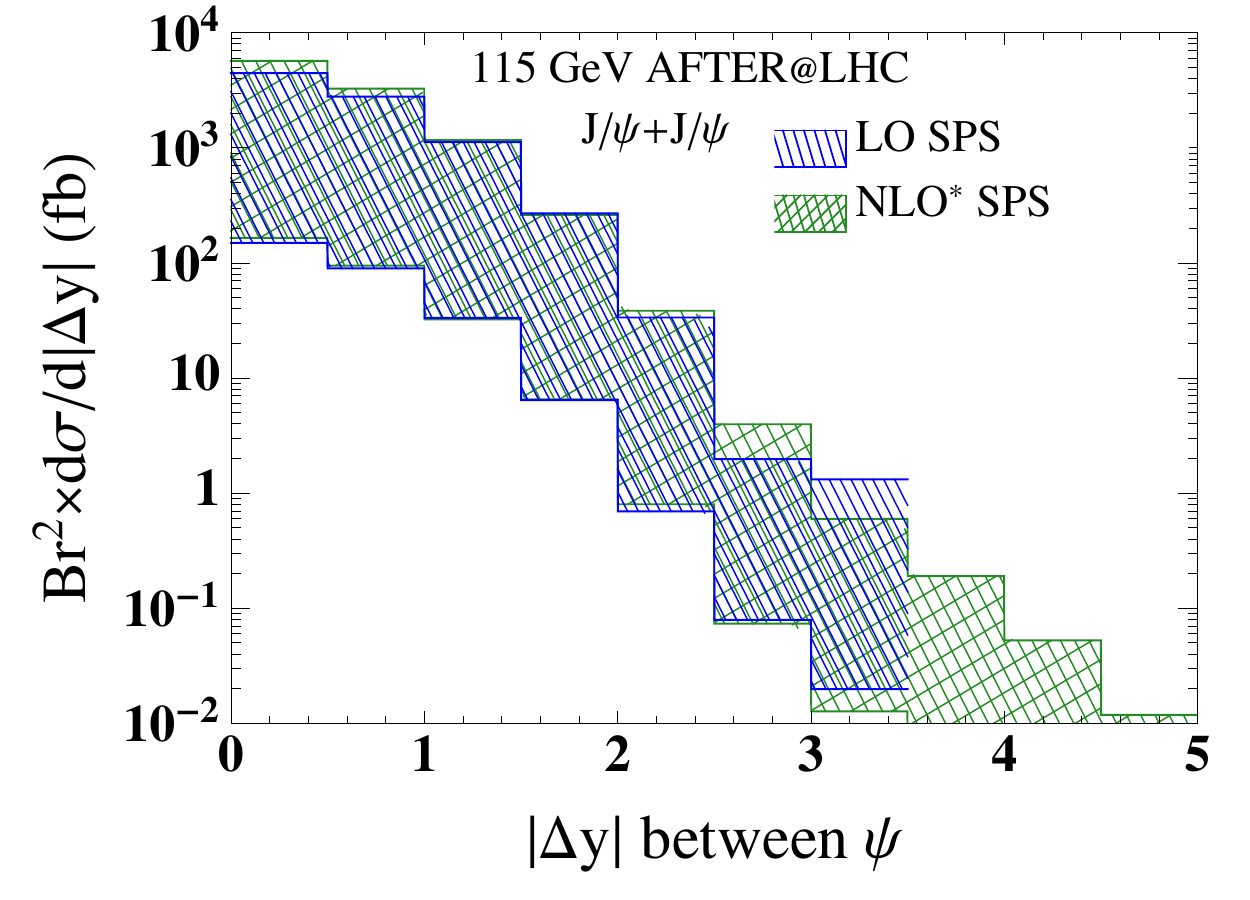}\label{fig:dsiglovsnlob}}
\subfloat[Pair Invariant mass]{\includegraphics[width=0.49\textwidth,draft=false]{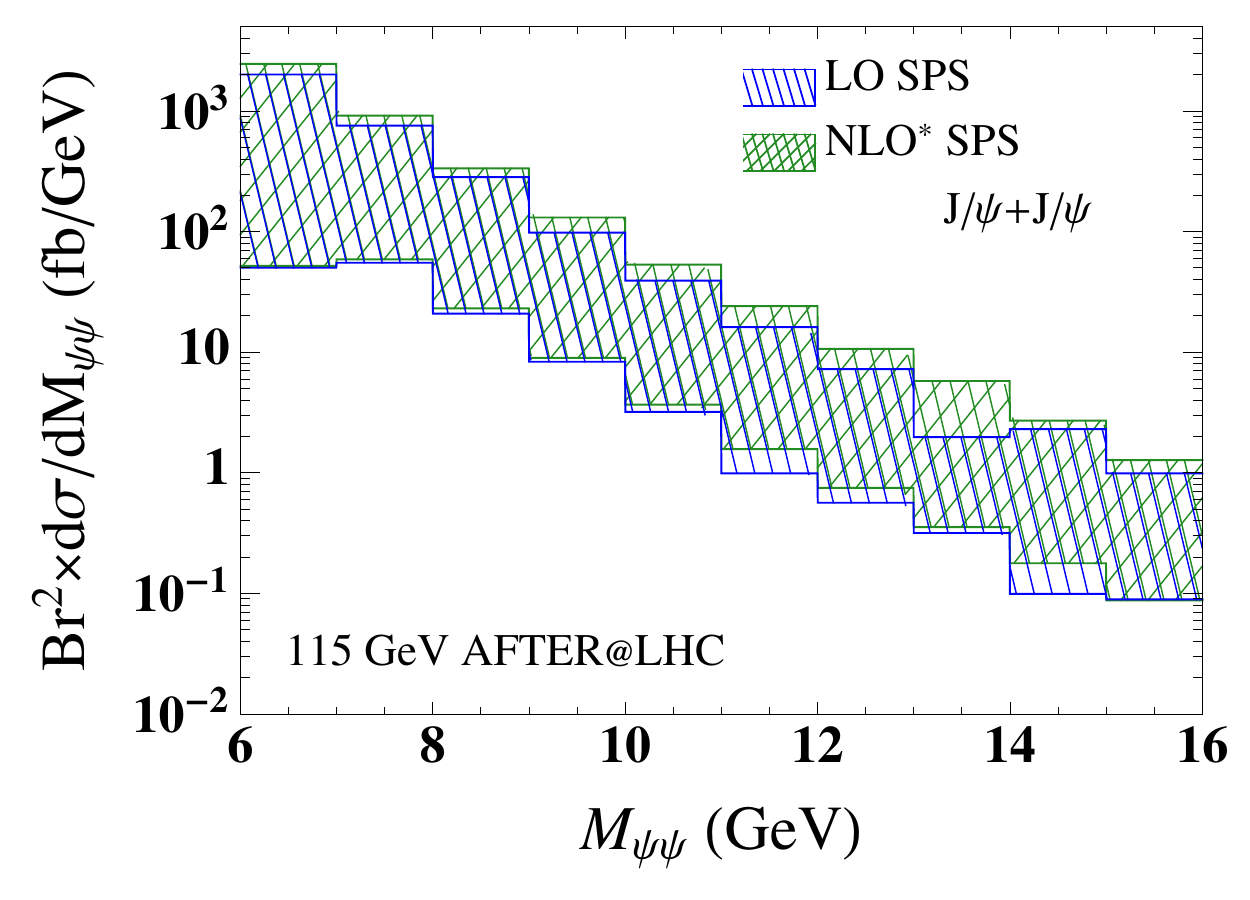}\label{fig:dsiglovsnloc}}\\
\subfloat[Leading $P_T$ among the $J/\psi$ pair]{\includegraphics[width=0.49\textwidth,draft=false]{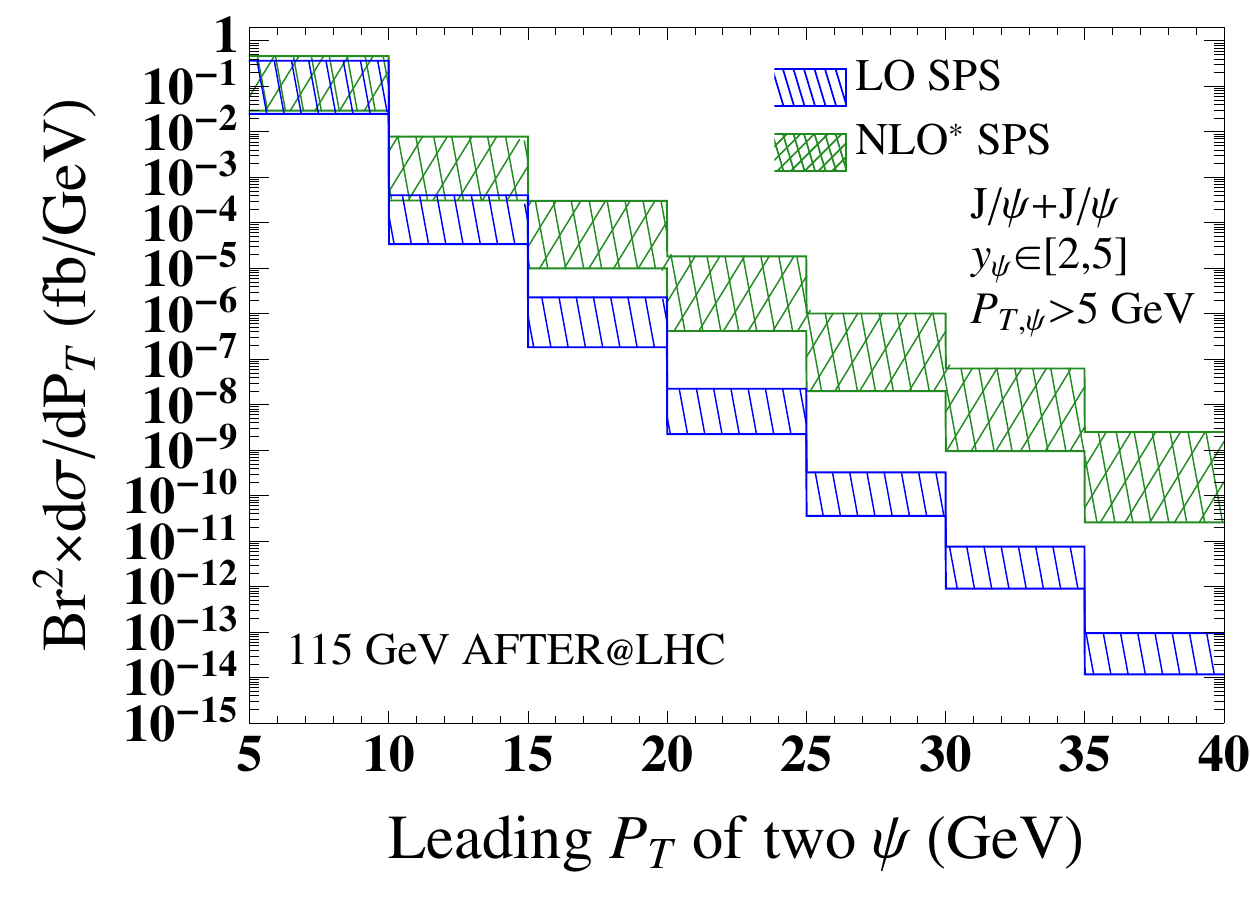}\label{fig:dsiglovsnlod}}
\subfloat[Pair transverse momentum]{\includegraphics[width=0.49\textwidth,draft=false]{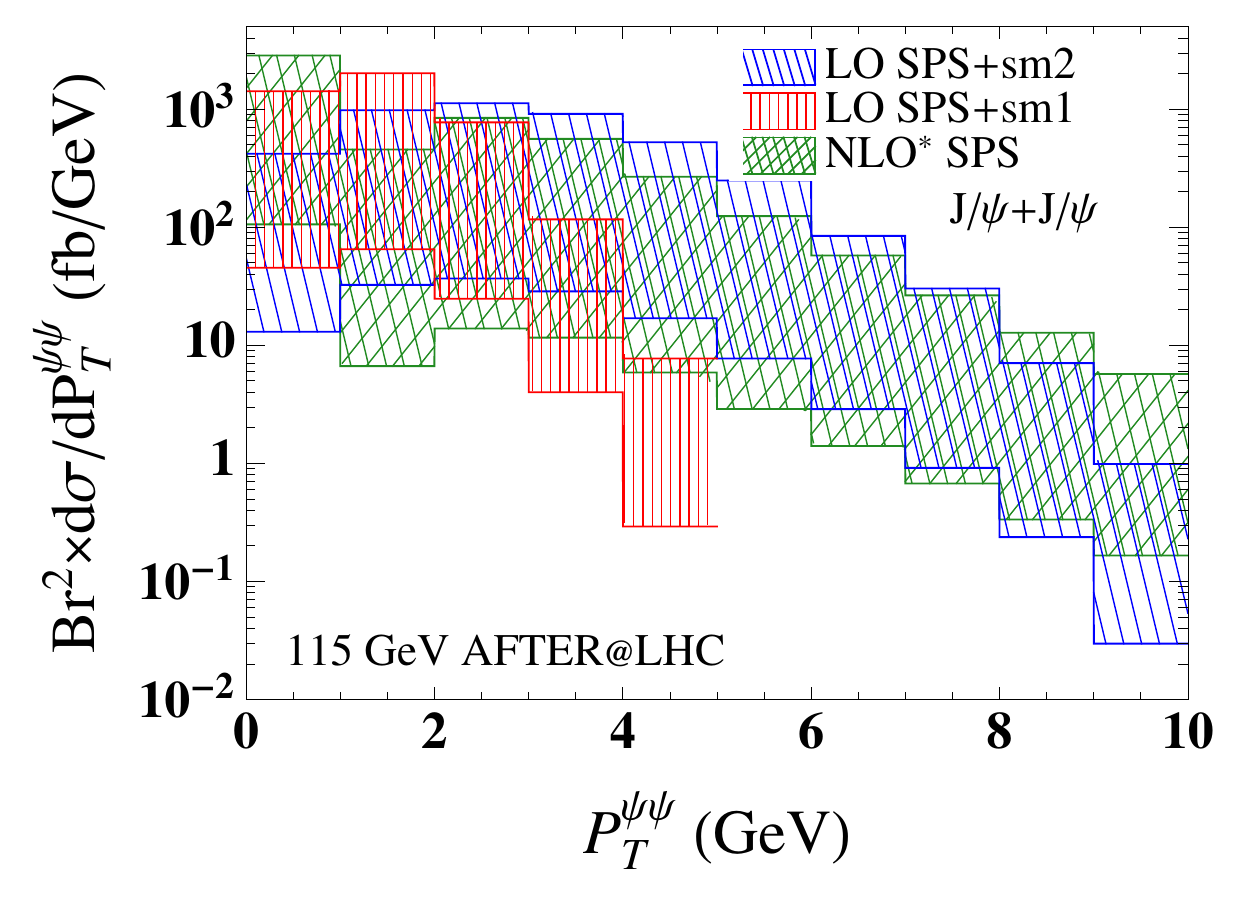}\label{fig:dsiglovsnloa}}
\caption{LO vs. NLO$^{\star}$ differential distributions.
}
\label{fig:dsiglovsnlo}
\end{center}\vspace*{-1cm}
\end{figure}

Figs.~\ref{fig:dsiglovsnlo} show the comparison between LO results and NLO$^\star$ results (which are known to reproduce well the full NLO~\cite{Sun:2014gca}). The invariant-mass and rapidity-difference spectra are not affected by the real emission at $\alpha_S^5$ . Indeed, in the low-$P_T$ region, the Born topologies are dominant, and there is no kinematical enhancement in the real-emission topologies which could compensate the $\alpha_S$ suppression. Only when one goes to large transverse momenta, these are enhanced and can become dominant. This explains the  difference in the slope as a function of the leading $P_T$ in \cf{fig:dsiglovsnlod}. The results are however similar for $P_T<10$ GeV where the cross sections are larger than 0.1 fb.

In addition, as we already discussed in Ref.\cite{Lansberg:2014swa}, at LO, a $2\to 2$ kinematics for SPS would result in a transverse momentum of the $J/\psi$-pair $P_T^{\psi\psi}$ being zero and in a trivial LO  distribution on \cf{fig:dsiglovsnloa}.
 This is however not the case if one takes into account a possible intrinsic $k_T$ of the initial partons which can also been considered as a part of QCD radiative corrections -- initial-state radiations to be precise. Such a smearing can be phenomenologically be accounted for and compared to a pQCD result. To do so, we have smeared the kinematics of LO events using a Gaussian distribution with $\langle k_T \rangle=1\ \&\ 2$ GeV as  done in Refs.~\cite{Lansberg:2013qka,Lansberg:2014swa}. We stress that the value of $\langle k_T \rangle$ is essentially empirical, hence the choice of two values for illustration (resp. curves labelled sm1 and sm2). This can thus be compared with our NLO$^\star$ curves in the accessible domain with ${\cal O} (20) $fb$^{-1}$ at AFTER@LHC, that is $P_T^{\psi\psi}<10$ GeV. One sees that the smearing mimics relatively well the effect of the QCD corrections with $\langle k_T \rangle=2$ GeV which we will use in the following for the comparison with the DPS yield. Overall, the $P_T^{\psi\psi}$ distribution is obviously very different than a single peak at 0. 

\section{Predictions at AFTER@LHC}

We are now in the position to present our numerical results at $\sqrt{s}=115$ GeV in $p+p$ collisions. The total cross section we obtained  are given in \ct{xsectionscc}, \ref{xsectionscb} and \ref{xsectionsbb}. The results have been multiplied by the branching ratios into a  muon pair and they are all in unit of fb. In general, we have
\bq
\sigma^{\Upsilon\Upsilon\to 4\mu}\ll \sigma^{\psi\Upsilon \to 4\mu}\ll \sigma^{\psi\psi\to 4\mu}.
\eq
The DPS contributions decrease quickly when the mass threshold $M_{\Q_1}+M_{\Q_2}$ increases because of its square dependence of the initial-state parton luminosity. With the nominal integrated luminosity of $20$ fb$^{-1}$ proposed to be collected at AFTER@LHC, we find that the measurement double-bottomonium production is out of reach\footnote{We note that such a measurement has never been done anywhere else.} and one may be able to record a few $J/\psi+\Upsilon(1S)$ events, which receives substantial DPS contributions.  

\begin{table}[!hbtp] 
\begin{center}
\begingroup
\renewcommand{\arraystretch}{1.3}
\begin{tabular}{c|ccc}\footnotesize
          &$J/\psi+J/\psi$ &$J/\psi+\psi(2S)$ & $\psi(2S)+\psi(2S)$ \\
\hline\hline
$\sigma_{\rm DPS}$ & $590^{+730}_{-210}$  & $19^{+23}_{-6.7}$ & $0.15^{+0.18}_{-0.052}$\\
$\sigma^{\rm CSM}_{\rm SPS}$  &  $700^{+3600}_{-560}$ & $85^{+440}_{-68}$ & $2.5^{+13}_{-2.0}$ \\
\end{tabular}
\caption{$\sigma(pp\to \Q_1+\Q_2+X) \times {\cal B}(\Q_1\to\mu^+\mu^-)\, {\cal B}(\Q_2\to\mu^+\mu^-)$ in units of fb at $\sqrt{s}=115$ GeV, where $\Q_1,\Q_2=J/\psi,\psi(2S)$. The DPS uncertainties are from  $\sigma_{\rm eff}$ 
and the SPS ones from $m_Q$ and the scales.}
\label{xsectionscc}
\endgroup
\end{center}
\end{table}

\begin{table}[!hbtp] 
\begin{center}
\begingroup
\renewcommand{\arraystretch}{1.3}
\begin{tabular}{c|ccc}\footnotesize
          &$J/\psi+\Upsilon(1S)$ &$J/\psi+\Upsilon(2S)$ & $J/\psi+\Upsilon(3S)$ \\
\hline
$\sigma_{\rm DPS}$ & $0.17^{+0.21}_{-0.058}$  & $0.037^{+0.045}_{-0.013}$ & $0.018^{+0.023}_{-0.0063}$ \\
$\sigma^{\rm NRQCD}_{\rm SPS}$  &  $<0.69$ & $<0.14$ & $<0.11$ \\
\hline\hline
& $\psi(2S)+\Upsilon(1S)$ & $\psi(2S)+\Upsilon(2S)$ & $\psi(2S)+\Upsilon(3S)$\\
\hline
$\sigma_{\rm DPS}$ &$2.6\cdot 10^{-3}~^{+3.2\cdot 10^{-3}}_{-9.1\cdot 10^{-4}}$ & $5.7\cdot 10^{-4}~^{+6.9\cdot 10^{-4}}_{-2.0\cdot 10^{-4}}$ & $2.8\cdot 10^{-4}~^{+3.4\cdot 10^{-4}}_{-9.8\cdot 10^{-5}}$
\\
$\sigma^{\rm NRQCD}_{\rm SPS}$  & $<0.031$ & $<5.4\cdot 10^{-3}$ & $<3.0\cdot 10^{-3}$\\
\end{tabular}
\caption{$\sigma(pp\to \Q_1+\Q_2+X) \times {\cal B}(\Q_1\to\mu^+\mu^-){\cal B}(\Q_2\to\mu^+\mu^-)$ in units of fb with $\sqrt{s}=115$ GeV, where $\Q_1=J/\psi,\psi(2S)$ and $\Q_2=\Upsilon(1S),\Upsilon(2S),\Upsilon(3S)$. For SPS production,  only the upper limits of the yields are given (see text). The DPS uncertainties are from  $\sigma_{\rm eff}$.}
\label{xsectionscb}
\endgroup
\end{center}
\end{table}

One should however always keep in mind that $\sigma_{\rm SPS}$ for $\psi+\Upsilon$ production strongly depends on the CO LDMEs. We have investigated this dependence  in \ref{appA2} with four different sets of LDMEs and the results vary up to one order of magnitude which precludes any strong conclusions\footnote{For convenience and possible future studies, we have tabulated in \ref{appA1}  the values of all the relevant short-distance coefficients which can then be combined with any LDME set.}.
In addition, these LDMEs are usually fit 
from the experimental data at high transverse momentum region and are known to overestimate the single-quarkonium yields at low $P_T$ (see~\cite{Feng:2015cba} and references therein). This is also probably the case for quarkonium-pair production especially when they come from single gluon splittings. We have therefore find it only meaningful to show upper limits on $\sigma_{\rm SPS}$ for $\psi+\Upsilon$ production in Table.~\ref{xsectionscb}. These numbers are in any case at the limit of observability.

The quoted theoretical uncertainties in the tables result from the variation of  $\sigma_{\rm eff}$ within $5\pm 2.75$ mb for the DPS yields and from the  scale uncertainties as well as heavy-quark-mass uncertainties for the SPS yields, as discussed in Sec.\ref{sec:meth}.

As regards double-charmonium production, about 10 thousand events could be collected per year --which is more than what has so far been collected by LHCb and CMS.
In the analysis of the differential distributions, we therefore only focus on these and, in particular, on  $J/\psi$-pair production. We show three interesting distributions without kinematical cuts. Along the lines of~\cite{Massacrier:2015qba},  we also used the LHCb kinematical acceptance, \ie\ the rapidity of $J/\psi$ restricted to be in the interval of $[2,5]$.

\begin{table}[!hbtp] 
\begin{center}
\begingroup
\renewcommand{\arraystretch}{1.3}
\begin{tabular}{c|ccc}\footnotesize
          &$\Upsilon(1S)+\Upsilon(1S)$ &$\Upsilon(2S)+\Upsilon(2S)$ & $\Upsilon(3S)+\Upsilon(3S)$  \\
\hline
$\sigma_{\rm DPS}$ & $1.2\cdot 10^{-5}~^{+1.4\cdot 10^{-5}}_{-4.0\cdot 10^{-6}}$  & $5.6\cdot 10^{-7}~^{+6.8\cdot 10^{-7}}_{-1.9\cdot 10^{-7}}$ & $1.4\cdot 10^{-7}~^{+1.7\cdot 10^{-7}}_{-4.7\cdot 10^{-8}}$ 
\\
$\sigma^{\rm CSM}_{\rm SPS}$  &  $2.8\cdot10^{-3}~^{+1.3\cdot 10^{-2}}_{-2.2\cdot 10^{-3}}$ & $3.5\cdot10^{-4}~^{+1.7\cdot 10^{-3}}_{-2.8\cdot 10^{-4}}$ & $2.2\cdot 10^{-4}~^{+1.1\cdot 10^{-3}}_{-1.8\cdot 10^{-4}}$ \\
\hline\hline
                 &$\Upsilon(1S)+\Upsilon(2S)$ & $\Upsilon(1S)+\Upsilon(3S)$                           & $\Upsilon(2S)+\Upsilon(3S)$ \\
\hline
$\sigma_{\rm DPS}$                              & $5.1\cdot 10^{-6}~^{+6.2\cdot 10^{-6}}_{-1.7\cdot 10^{-6}}$ & $2.5\cdot 10^{-6}~^{+3.0\cdot 10^{-6}}_{-8.7\cdot 10^{-7}}$ & $5.5\cdot 10^{-7}~^{+6.7\cdot 10^{-7}}_{-1.9\cdot 10^{-7}}$
\\
$\sigma^{\rm CSM}_{\rm SPS}$  & $2.0\cdot 10^{-3}~^{+9.3\cdot 10^{-3}}_{-1.6\cdot 10^{-3}}$ & $1.6\cdot 10^{-3}~^{+7.4\cdot 10^{-3}}_{-1.3\cdot 10^{-3}}$ & $5.6\cdot 10^{-4}~^{+2.6\cdot 10^{-3}}_{-4.4\cdot 10^{-4}}$\\
\end{tabular}
\caption{$\sigma(pp\to \Q_1+\Q_2+X) \times {\cal B}(\Q_1\to\mu^+\mu^-){\cal B}(\Q_2\to\mu^+\mu^-)$ in units of fb with $\sqrt{s}=115$ GeV, where $\Q_1,\Q_2=\Upsilon(1S),\Upsilon(2S),\Upsilon(3S)$. The DPS uncertainties are from  $\sigma_{\rm eff}$ 
and the SPS ones from the $m_Q$ and the scales.}
\label{xsectionsbb}
\endgroup
\end{center}
\end{table}

\begin{figure}[hbt!]
\begin{center}
\subfloat{\includegraphics[width=0.49\textwidth,draft=false]{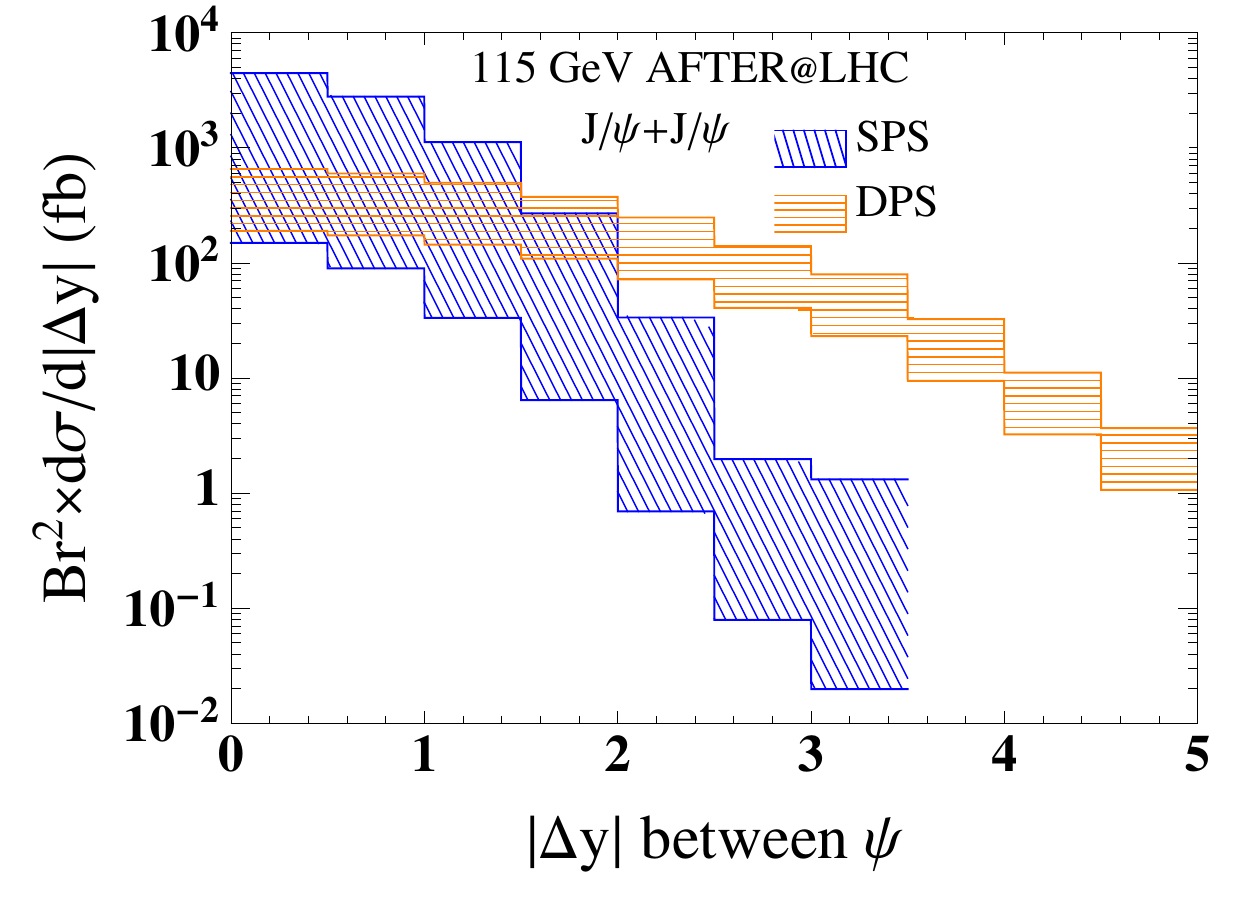}\label{fig:dsigb}}
\subfloat{\includegraphics[width=0.49\textwidth,draft=false]{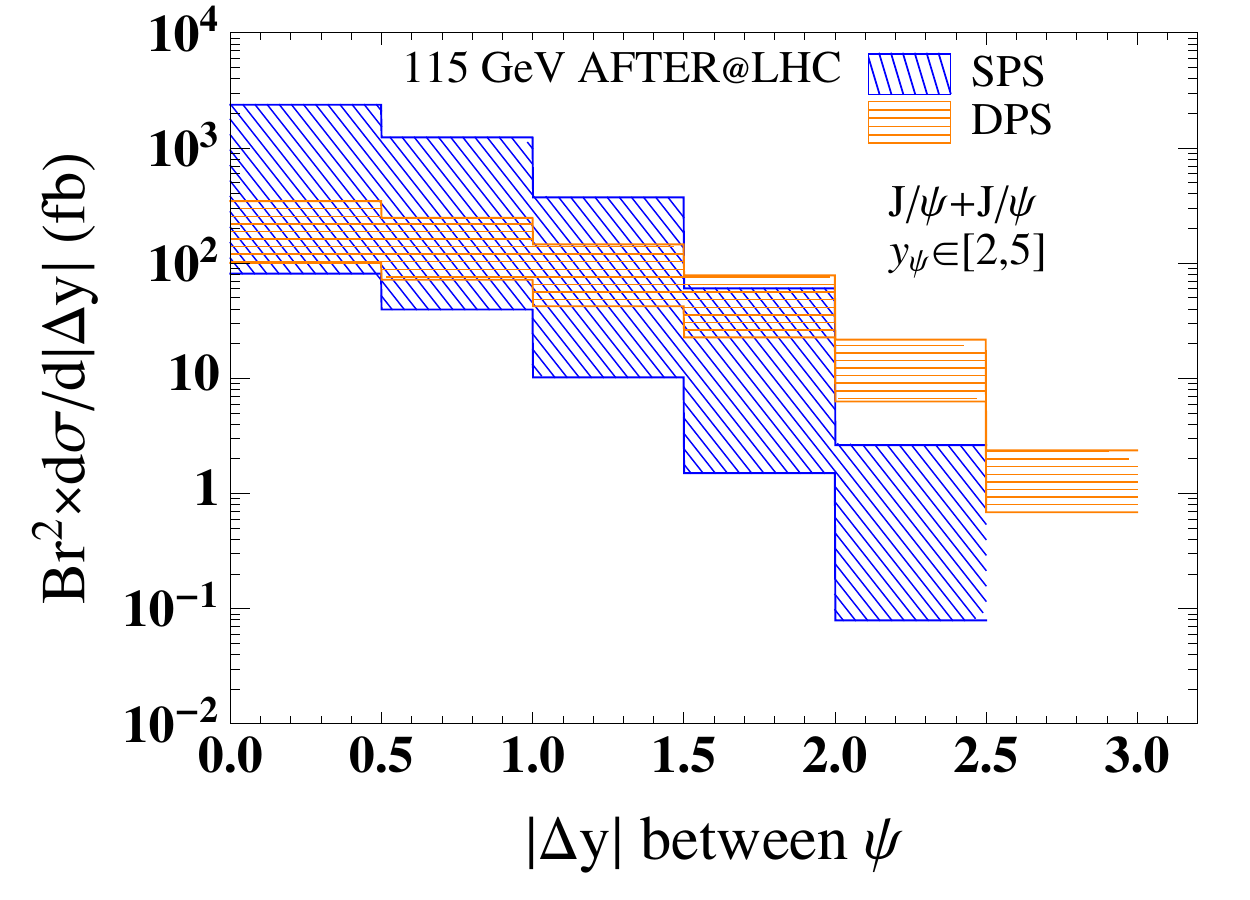}\label{fig:dsigLHCbb}}
\caption{Differential cross section as a function of the absolute rapidity difference of the $J/\psi$ pair, without (left) or with
(right) a rapidity cut.}
\label{fig:dsigdDeltay}
\end{center}\vspace*{-1cm}
\end{figure}

\begin{figure}[hbt!]
\begin{center}
\subfloat{\includegraphics[width=0.49\textwidth,draft=false]{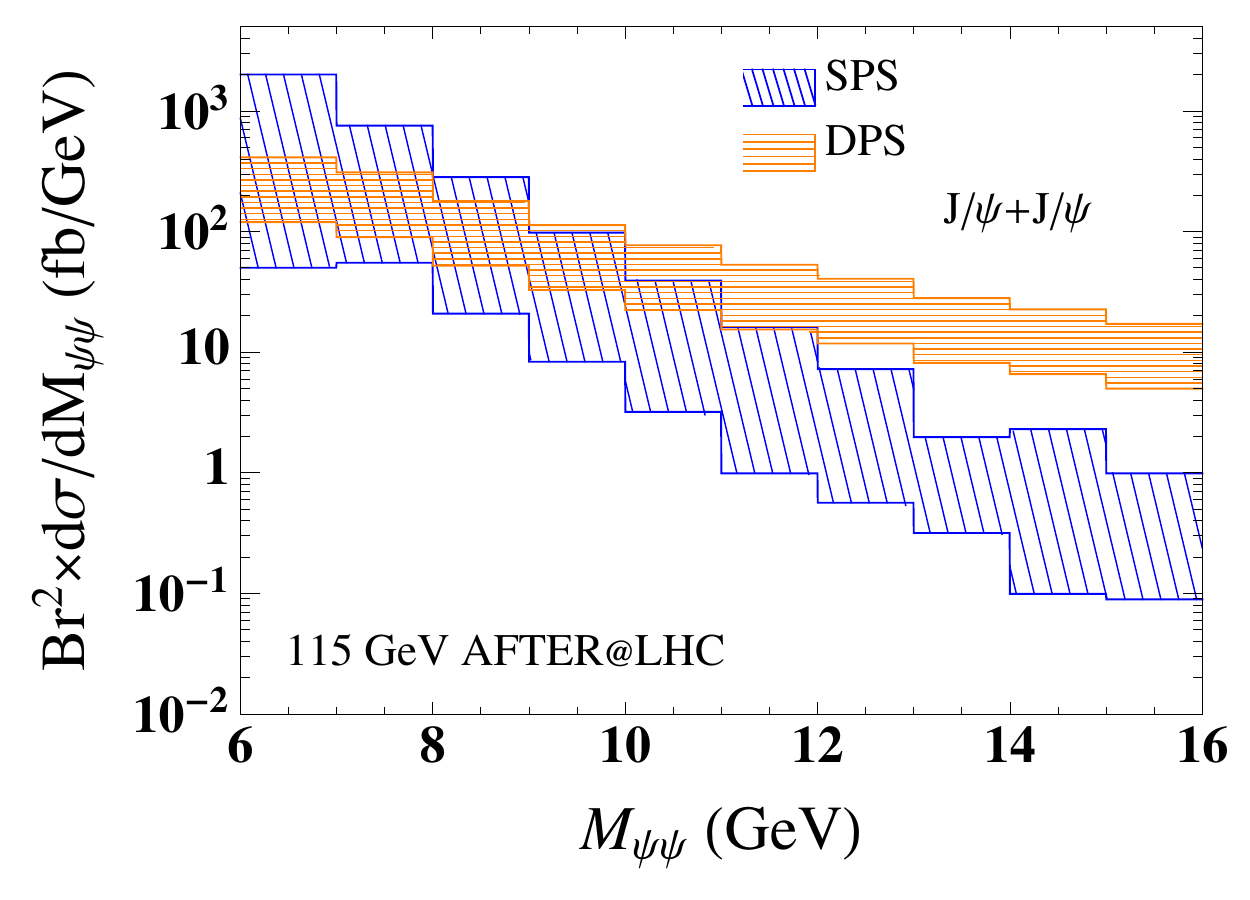}\label{fig:dsigc}}
\subfloat{\includegraphics[width=0.49\textwidth,draft=false]{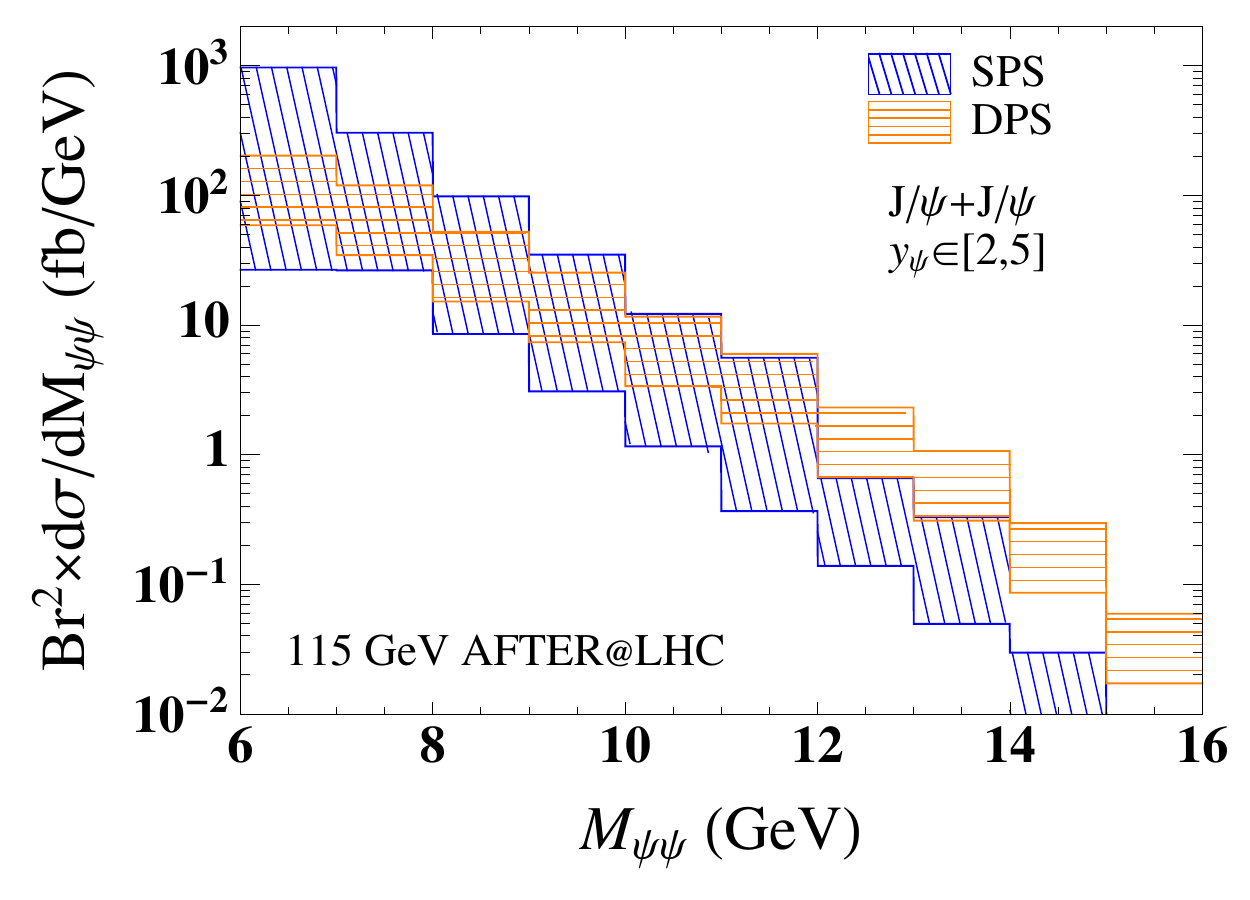}\label{fig:dsigLHCbc}}
\caption{Differential cross section as a function of the invariant mass of the $J/\psi$ pair, without (left) or with
(right) a rapidity cut.}
\label{fig:dsigdM}
\end{center}\vspace*{-1cm}
\end{figure}

\begin{figure}[hbt!]
\begin{center}
\subfloat{\includegraphics[width=0.49\textwidth,draft=false]{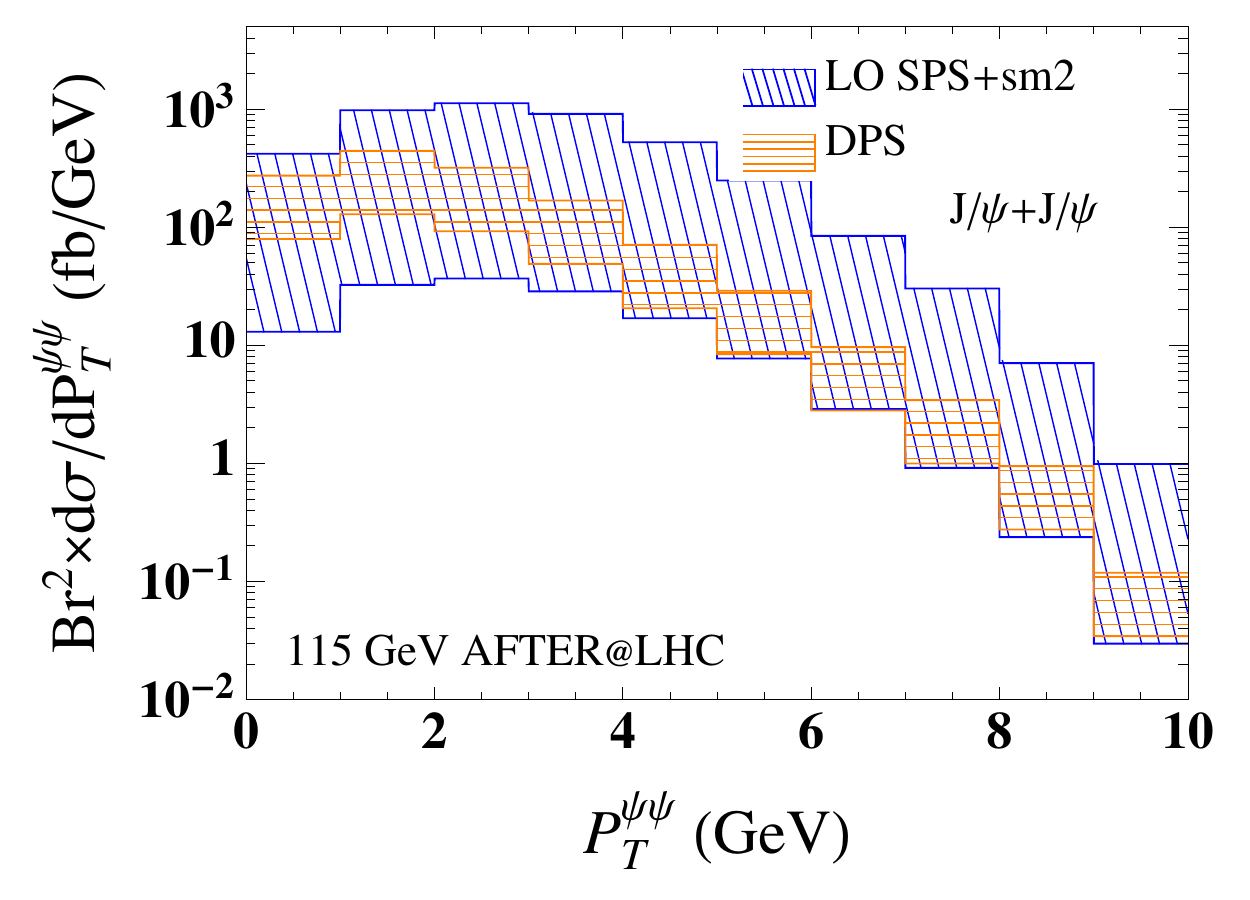}\label{fig:dsiga}}
\subfloat{\includegraphics[width=0.49\textwidth,draft=false]{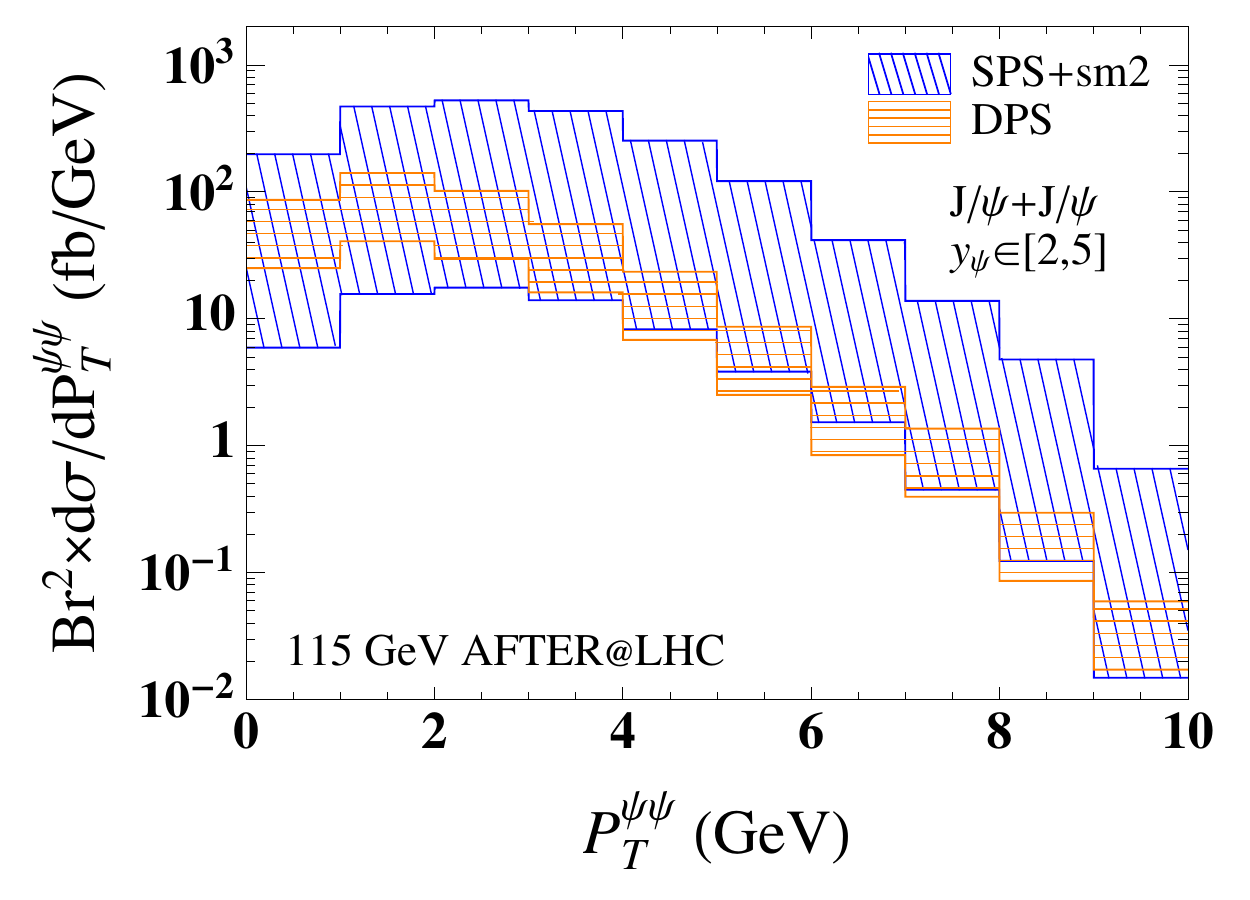}\label{fig:dsigLHCba}}
\caption{Differential cross section as a function of the transverse momentum of the $J/\psi$ pair, without (left) or with
(right) a rapidity cut.}
\label{fig:dsigdPtpsipsi}
\end{center}\vspace*{-1cm}
\end{figure}

\begin{figure}[hbt!]
\begin{center}
\subfloat{\includegraphics[width=0.49\textwidth,draft=false]{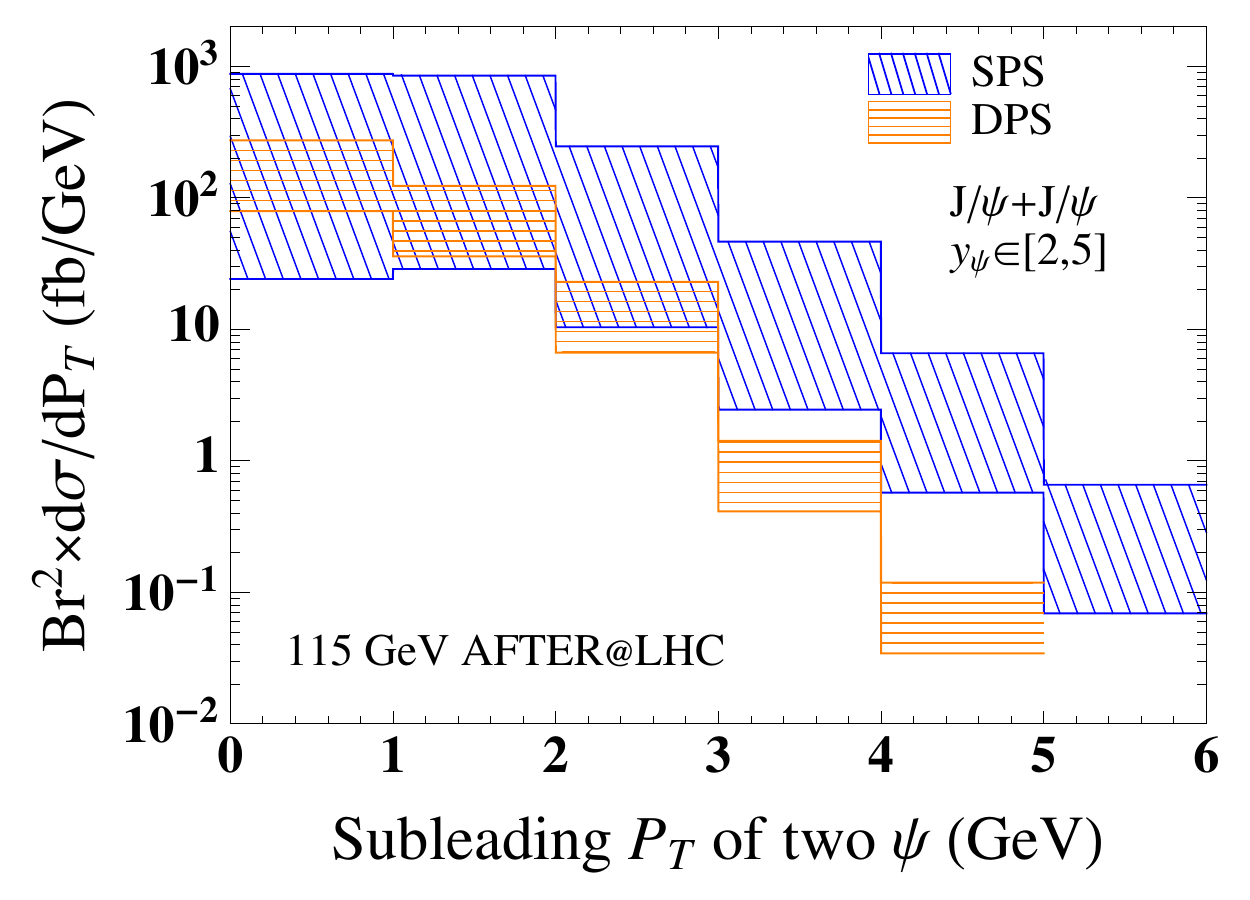}\label{fig:dsigLHCbd}}
\subfloat{\includegraphics[width=0.49\textwidth,draft=false]{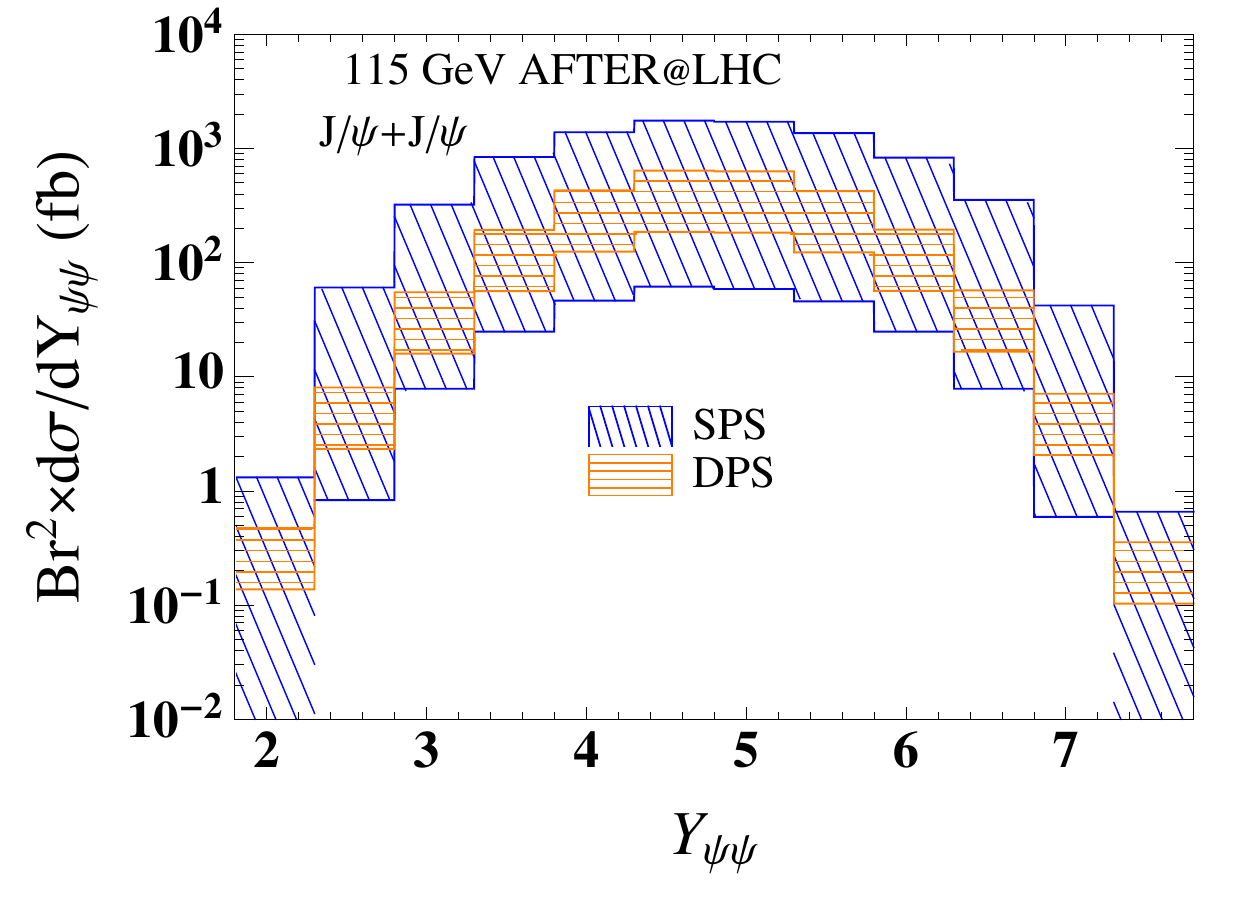}\label{fig:dsigd}}
\caption{Differential cross section as  a function of (left) the sub-leading $P_T$ with a rapidity cut  and (right)the rapidity of the $J/\psi$ pair.}
\label{fig:dsigdYanddpt}
\end{center}\vspace*{-1cm}
\end{figure}

The absolute rapidity difference between the $J/\psi$ pair is expected to be a good observable to discriminate the DPS and SPS contributions. This was first pointed out in Ref.~\cite{Kom:2011bd} and this was used later on  by D0 collaboration~\cite{Abazov:2014qba} to extract $\sigma_{\rm eff}$ from double-$J/\psi$ production at the Tevatron. The DPS events should have a broader distribution in $\Delta y$ than the SPS ones, because two (relatively) independent hard interactions happen simultaneously in DPS while the two $J/\psi$ from SPS are more correlated.  The situation still does not change at AFTER@LHC without or with cut as \cf{fig:dsigdDeltay} (left) and (right) show. In the latter case, the restriction to negative rapidities in the centre-of-mass obviously reduce the $\Delta y$ range. Starting from $\Delta y=2$, the DPS events dominate the SPS events. A ratio DPS/SPS of 10 is obtained for $\Delta y>2$. The distribution of the invariant mass for the $J/\psi$ pair $M_{\psi\psi}$ reflects a similar information as the $\Delta y$ distribution. Hence, it follows that the $M_{\psi\psi}$ spectra of DPS are also broader than those of SPS, which can be seen on \cf{fig:dsigdM} (left) and (right).

As we discussed earlier, predictions for the $P_T^{\psi\psi}$ dependence of the SPS yield depend much on the $k_T$ smearing of the initial partons which can mimic a part of the QCD corrections. Due to the relative smaller yields at AFTER@LHC energies  than at LHC energies, one can only access  $P_T^{\psi\psi}<10$ GeV, as illustrated on \cf{fig:dsigdPtpsipsi}. In such a kinematical region, the $k_T$ smearing effect makes the SPS spectrum as broad as the DPS one with $\langle k_T \rangle = 2$ GeV.

Finally, we present on \cf{fig:dsigdYanddpt} the cross section as a function of the total rapidity of the $J/\psi$ pair (right), $Y_{\psi\psi}$, and of the sub-leading $P_T$ between the $J/\psi$ pair (left). One sees that the sub-leading $P_T$ spectrum may be measured up to  6 GeV with AFTER@LHC.  
As regards the rapidity distribution, its maximum is obviously located at $Y_{\rm cms}=0$, that is  $Y=4.8$ in the laboratory frame.  One sees that one can expect some counts down to $Y_{\psi\psi}\simeq 2.5$ where $x_F\simeq \frac{2M_{\psi\psi}}{\sqrt{s}} \sinh(Y_{\psi\psi}-4.8) \simeq -0.5$. This is precisely the kinematical region where double intrinsic $c \bar c$  coalescence contributes on average~\cite{Vogt:1995tf}. Any modulation in the pair-rapidity distribution 
would sign the presence of such a contribution.

Finally, we have investigated the impact of using different (double)PDFs (MSTW2008NLO~\cite{Martin:2009iq}, CTEQ6L1~\cite{Pumplin:2002vw}, GS09 dPDF~\cite{Gaunt:2009re}) on differential distributions are also shown in \cf{fig:dsigDPS}; they are found to be  moderate in all cases.

\begin{figure}[t!]
\begin{center}
\subfloat[Pair transverse momentum ]{\includegraphics[width=0.49\textwidth,draft=false]{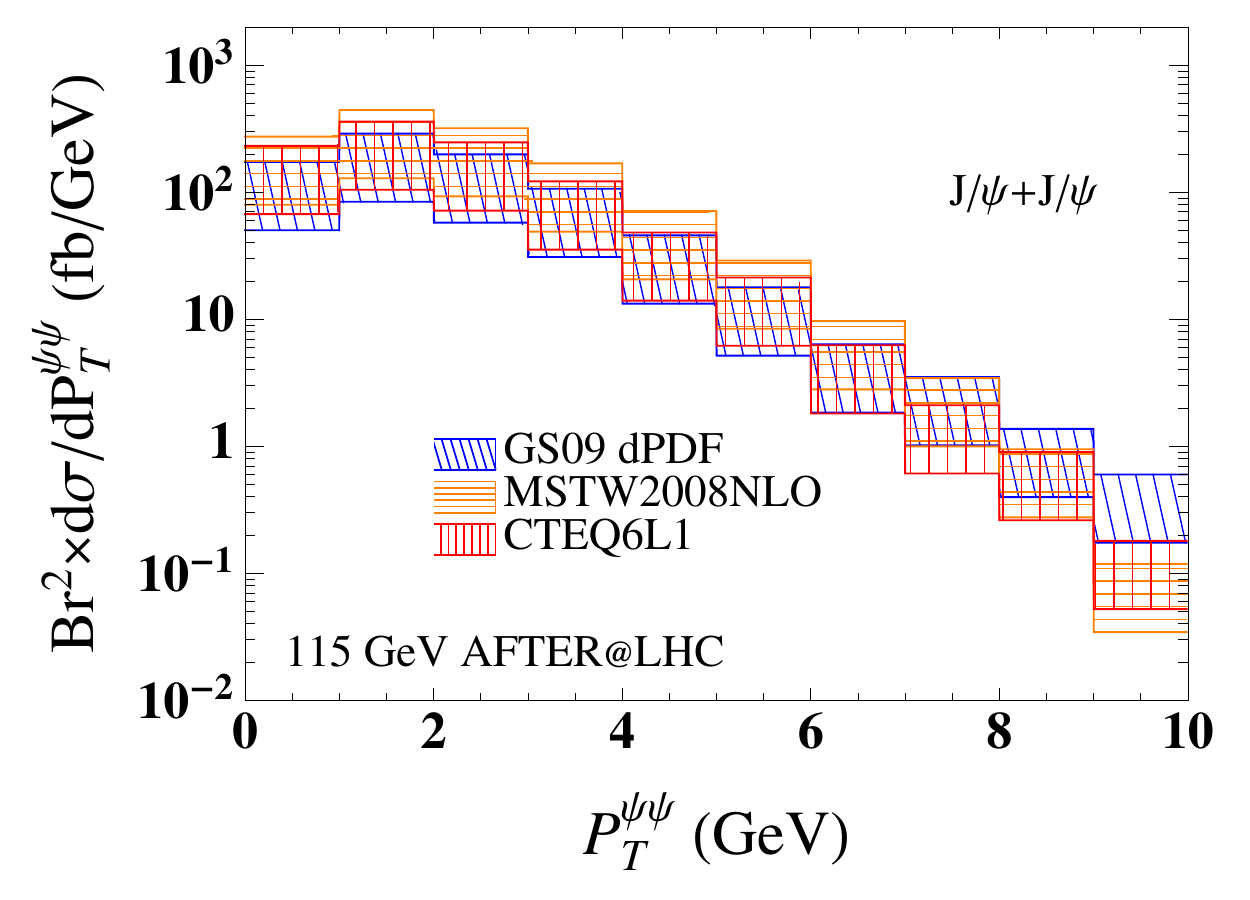}\label{fig:dsigDPSa}}
\subfloat[Absolute rapidity difference between both $J/\psi$]{\includegraphics[width=0.49\textwidth,draft=false]{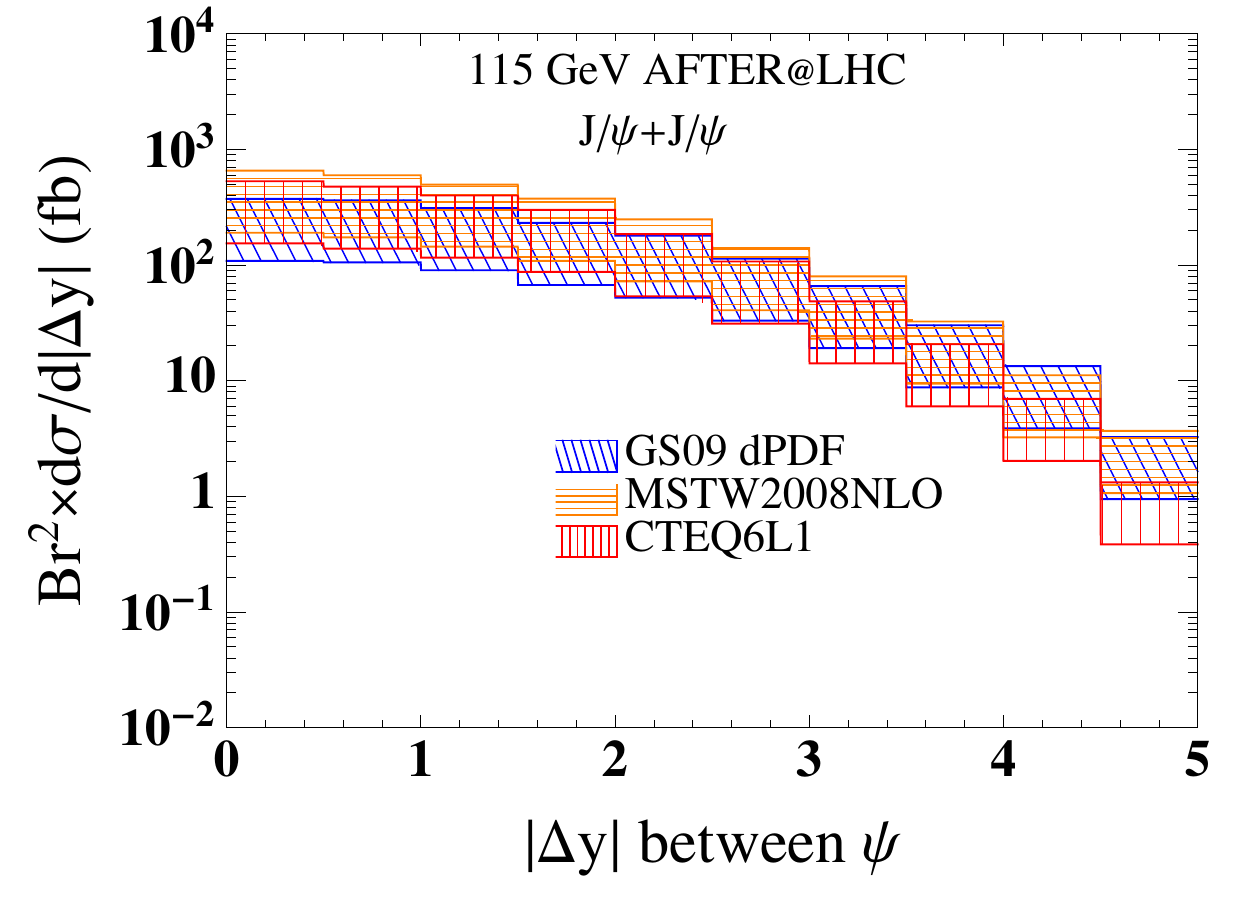}\label{fig:dsigDPSb}}\\
\subfloat[Pair invariant mass]{\includegraphics[width=0.49\textwidth,draft=false]{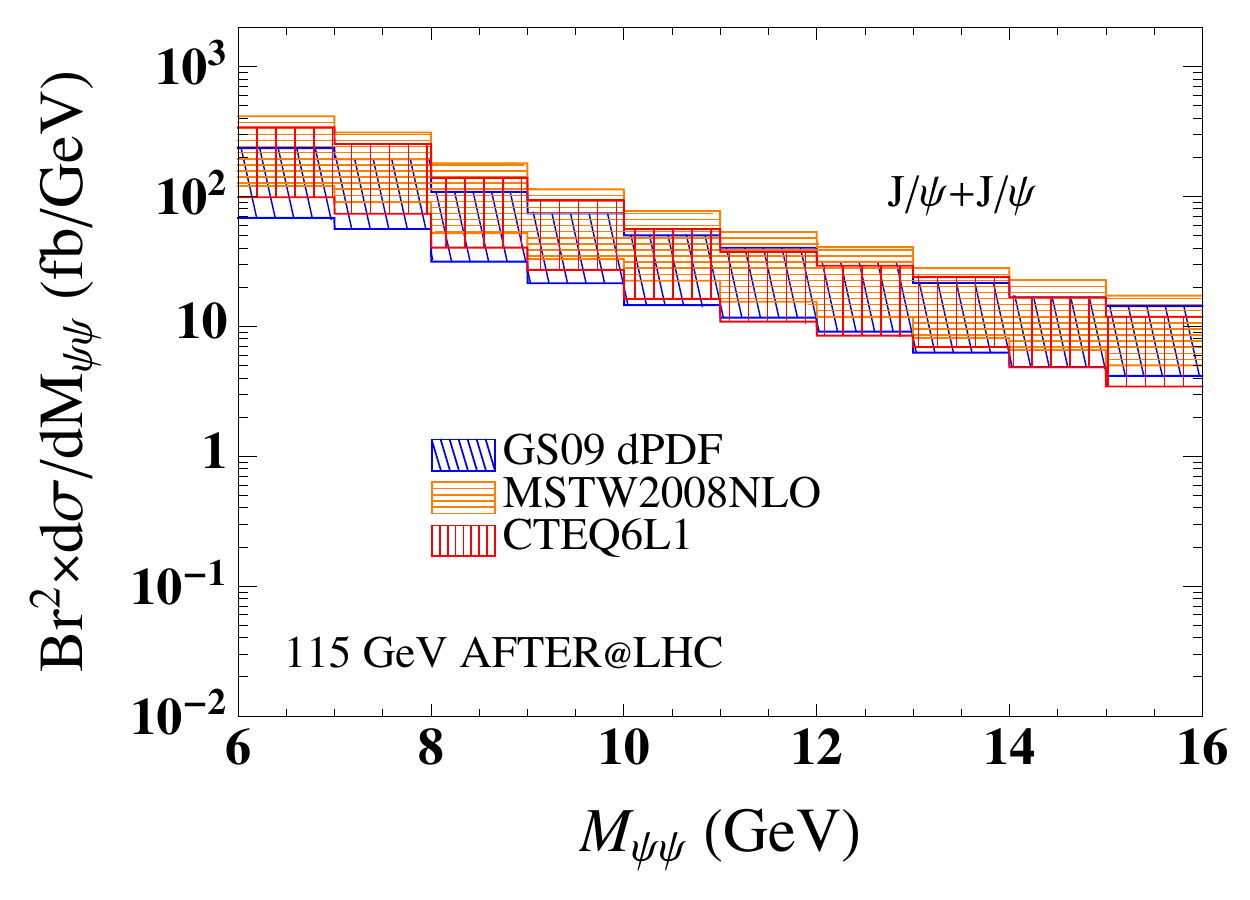}\label{fig:dsigDPSc}}
\subfloat[Pair rapidity]{\includegraphics[width=0.49\textwidth,draft=false]{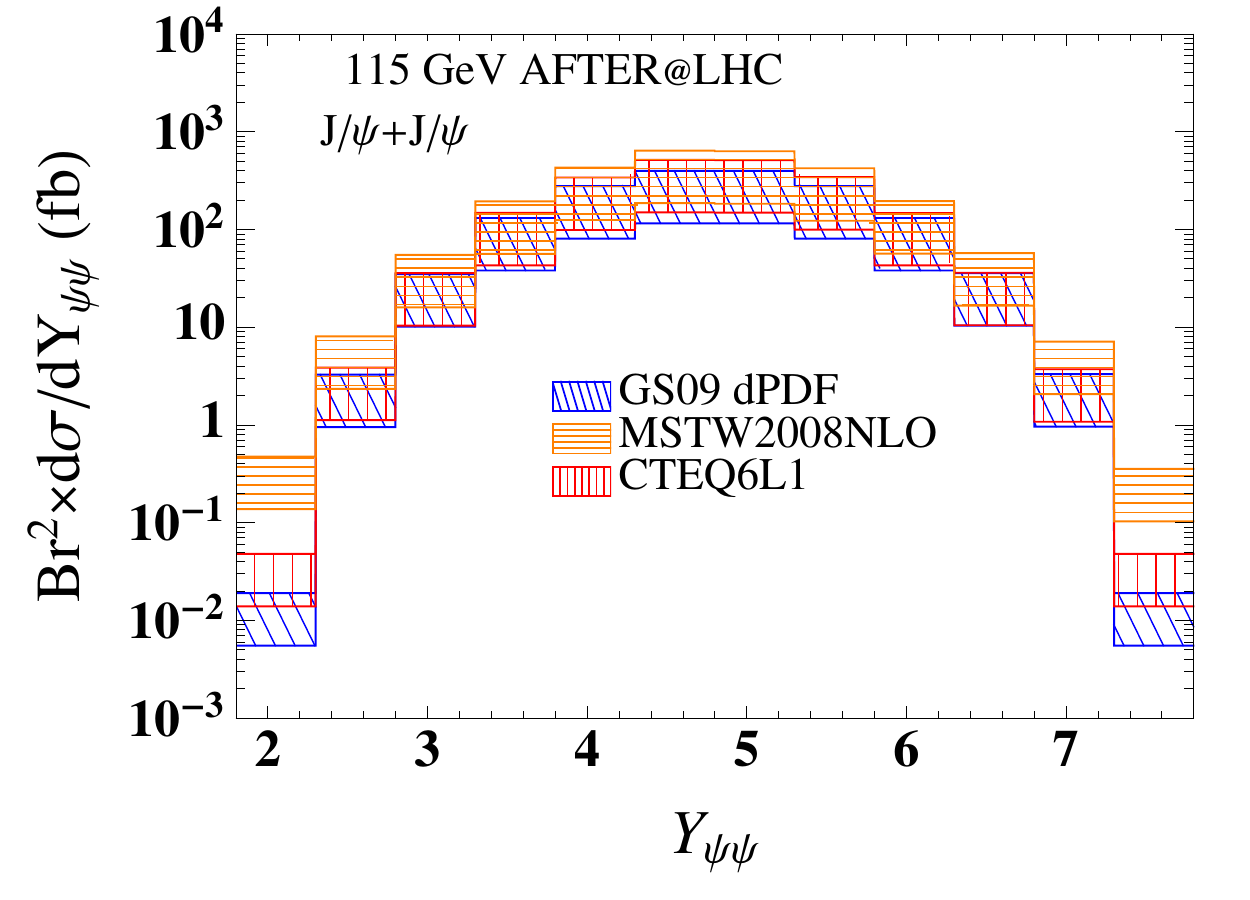}\label{fig:dsigDPSd}}
\caption{Differential distributions for DPS with various PDFs: (a) transverse momentum spectrum; (b) absolute rapidity difference ; (c) invariant mass distribution; (d) rapidity of $J/\psi$ pair.
}
\label{fig:dsigDPS}
\end{center}\vspace*{-1cm}
\end{figure}

\section{Conclusion}

We have discussed double-quarkonium production in proton-proton collisions at a fixed-target experiment using the LHC proton beams, AFTER@LHC. 
These processes have lately attracted much attention, both in the theorist and experimentalist communities. They are expected to be good 
observables to further constrain the various models describing heavy-quarkonium production. Double-quarkonium production also provides a good opportunity to study DPS since the yields of single quarkonium production is large and their decay to four muons is a clean signal at a hadron colliders. AFTER@LHC provides very appealing opportunities to study these observables with a LHCb-like detector and in new energy region. 

In this paper, we have studied both DPS and SPS contributions for double-quarkonium production. These processes include $\psi(n_1S)+\psi(n_2S)$, $\psi(n_1S)+\Upsilon(m_1S)$ and $\Upsilon(m_1S)+\Upsilon(m_2S)$ with $n_1,n_2=1,2$ and $m_1,m_2=1,2,3$. DPS contributions are estimated in a data-driven way, while SPS ones are calculated at LO in non-relativistic QCD (NRQCD)~\cite{Bodwin:1994jh}, more precisely in the CSM for $\psi(n_1S)+\psi(n_2S)$ and $\Upsilon(m_1S)+\Upsilon(m_2S)$ and accounting for CO contributions for $\psi(n_1S)+\Upsilon(m_1S)$. From our calculations, we find that ten thousand of double-charmonium events can indeed be measured at AFTER@LHC with the yearly integrated luminosity of $20$ fb$^{-1}$.  In the most backward region, a careful analysis of the rapidity distribution could also uncover double intrinsic $c \bar c$ coalescence contributions. In general, future measurements on double-charmonium production can provide extremely valuable information on QCD, in particular  important tests on the factorisation formula for DPS and the energy (in)dependence of $\sigma_{\rm eff}$.


\appendix

\section{Charmonium-bottomonium pair production in NRQCD\label{appA}}

\subsection{Short-distance coefficients for charmonium-bottomonium pair production\label{appA1}}

In NRQCD~\cite{Bodwin:1994jh}, the cross section for a charmonium $\C$ and a bottomonium $\B$ production can  systematically be written as
\bqa
\sigma(\C+\B)&=&\sum_{n_1,n_2}{\sigma(c\bar{c}[n_1]+b\bar{b}[n_2])\times\langle\mathcal{O}^{\C}(n_1) \rangle \times \langle\mathcal{O}^{\B}(n_2) \rangle},
\eqa
where $n_1,n_2$ are different possible Fock states, $\sigma(c\bar{c}[n_1]+b\bar{b}[n_2])$ is the short-distance coefficient (SDC) for the production of a charm-quark pair in the Fock state $n_1$ and a bottom-quark pair in the Fock state $n_2$ simultaneously. The LDMEs $\langle\mathcal{O}^{\C}(n_1) \rangle$ and $\langle\mathcal{O}^{\B}(n_2) \rangle$ should obey the velocity-scaling rules of NRQCD. In this appendix, we give the complete list of the SDCs for charmonium-bottomonium pair production at the leading order in $\alpha_s$ in proton-proton collisions at the center-of-mass energy of $\sqrt{s}=115$ GeV. This includes the contributions from $\ss,\so,\sps,\pj (J=0,1,2)$ for $S$-wave quarkonium production and from $\so,\pjs (J=0,1,2)$ for $P$-wave quarkonium production. There are in total $66$ non-vanishing channels to be computed. Such a computation is automatic in \HELACOnia~\cite{Shao:2012iz,Shao:2015vga}, but has never been carried out even at LHC energies. 
Thanks to the heavy-quark-spin symmetry of NRQCD, we have 
\bqa
\langle\mathcal{O}^{\C,\B}(\pj) \rangle=(2J+1)\times\langle\mathcal{O}^{\C,\B}(\p0) \rangle.
\eqa
We can thus define  new SDCs relevant for $\pj$
\bqa
\sigma(c\bar{c}[\sum_{J=0}^2{\pj}]+b\bar{b}[n_2])&\equiv&\sum_{J=0}^2{(2J+1)\times \sigma(c\bar{c}[{\pj}]+b\bar{b}[n_2])},\nonumber\\
\sigma(c\bar{c}[n_1]+b\bar{b}[\sum_{J=0}^2{\pj}])&\equiv&\sum_{J=0}^2{(2J+1)\times \sigma(c\bar{c}[n_1]+b\bar{b}[\pj])}.
\eqa
Therefore, we have
\bqa
\!\!\sum_{J=0}^2{\sigma(c\bar{c}[\pj]+b\bar{b}[n_2])\times \langle\mathcal{O}^{\C}(\pj) \rangle \langle\mathcal{O}^{\B}(n_2) \rangle}
=\sigma(c\bar{c}[\sum_{J=0}^2{\pj}]+b\bar{b}[n_2])\times\langle\mathcal{O}^{\C}(\p0) \rangle \langle\mathcal{O}^{\B}(n_2) \rangle,\nonumber\\
\!\!\sum_{J=0}^2{\sigma(c\bar{c}[n_1]+b\bar{b}[\pj])\times \langle\mathcal{O}^{\C}(n_1) \rangle \langle\mathcal{O}^{\B}(\pj) \rangle}
=\sigma(c\bar{c}[n_1]+b\bar{b}[\sum_{J=0}^2{\pj}])\times\langle\mathcal{O}^{\C}(n_1) \rangle \langle\mathcal{O}^{\B}(\p0) \rangle.\nonumber\\
\!\!
\eqa
We display the numerical values for the SDCs for these Fock states in Table.~\ref{tab:SDC} with CTEQ6L1~\cite{Pumplin:2002vw} as our PDF set.

\begin{table*}[!hbtp] 
\begin{center}\small
\begin{tabular}{{c|}*{7}{c}}\hline\hline
Fock state & $b\bar{b}[\ss]$ & $b\bar{b}[\so]$ & $b\bar{b}[\sps]$ & $b\bar{b}[\sum_{J=0}^2{\pj}]$ & $b\bar{b}[\tpzs]$ & $b\bar{b}[\tpos]$ & $b\bar{b}[\tpts]$\\\hline 
$c\bar{c}[\ss]$ & - & $13^{+63}_{-10}$ & - & - & - & - & - \\
$c\bar{c}[\so]$ & $40^{+200}_{-32}$ & $770^{+4000}_{-620}$ & $2700^{+14000}_{-2200}$ & $720^{+4200}_{-590}$ & $160^{+950}_{-130}$ & $7.3^{+44.0}_{-6.0}$ & $43^{+250}_{-36}$ \\
$c\bar{c}[\sps]$ & - & $220^{+1100}_{-170}$ & $650^{+3500}_{-520}$ & $180^{+1100}_{-150}$ & $46^{+280}_{-38}$ & $2.0^{+12}_{-1.6}$ & $9.1^{+56}_{-7.6}$ \\
$c\bar{c}[\sum_{J=0}^2{\pj}]$ & - & $470^{+2700}_{-380}$ & $1200^{+7500}_{-990}$ & $330^{+2400}_{-280}$ & $31^{+220}_{-26.0}$ & $1.2^{+8.6}_{-1.}$ & $8.^{+58}_{-6.8}$ \\
$c\bar{c}[\tpzs]$ & - & $31^{+180}_{-25}$ & $210^{+1300}_{-180}$ & $25^{+180}_{-21}$ & $12^{+87}_{-10}$ & $0.37^{+2.6}_{-0.31}$ & $3.1^{+23}_{-2.7}$ \\
$c\bar{c}[\tpos]$ & - & $21^{+130}_{-18}$ & $69^{+430}_{-57}$ & $7.5^{+54}_{-6.4}$ & $3.6^{+26}_{-3.1}$ & $0.33^{+2.3}_{-0.28}$ & $1.0^{+7.2}_{-0.86}$ \\
$c\bar{c}[\tpts]$ & - & $21^{+120}_{-18}$& $7.5^{+310}_{-40}$ & $6.1^{+44}_{-5.2}$ & $3.0^{+22}_{-2.5}$ & $0.15^{+1.1}_{-0.13}$ & $0.79^{+5.8}_{-0.68}$ \\
\hline\hline
\end{tabular}
\end{center}
\caption{The SDCs (at the leading order in $\alpha_s$) for the various combinations of the Fock states contributing to charmonium-bottomonium pair production at $\sqrt{s}=115$ GeV. The unit of the SDCs of $c\bar{c}[n_1]+b\bar{b}[n_2]$ is fb/GeV$^{6+2L_1+2L_2}$, where $L_i=0$ when $n_i$ is $S$-wave and $L_i=1$ when $n_i$ is $P$-wave. The uncertainty quoted is coming from the variation of $\mu_F=\mu_R \in [\frac{1}{2}\mu_0,2\mu_0]$ ($\mu_0=\sqrt{4(m_c+m_b)^2+P_T^2}$) and the uncertainties  on $m_c=1.5\pm0.1$~GeV and $m_b=4.75\pm 0.25$~GeV.}
\label{tab:SDC}
\end{table*}

\subsection{Cross sections for single-parton scattering\label{appA2}}

From the SDCs given in Table.~\ref{tab:SDC} and the LDMEs extracted from the experimental data, we are now able to estimate the cross sections of charmonium+bottomonium pair production at $\sqrt{s}=115$ GeV. The values of the LDMEs however  significantly differ depending on the different  experimental input data and the different fit setup. For example, the CO LDMEs of $J/\psi$ extracted from $pp$ data can be quite different with or without NLO QCD corrections. Here, we will discuss the results based on four sets of LDMEs for charmonia and bottomonia, which can be described as follows:
\begin{enumerate}
\item[Set I:] This set is based on the LDMEs of $J/\psi,\psi(2S)$ and $\chi_c$ presented in Ref.~\cite{Braaten:1999qk} and those of $\Upsilon(1S),\Upsilon(2S),\Upsilon(3S)$ and $\chi_{b}(1P),\chi_b(2P)$ presented in Ref.~\cite{Kramer:2001hh}. They are extracted from Tevatron data with SDCs at LO in $\alpha_s$. The LDMEs of $\chi_{b}(3P)$ have been set to zero in the fit of Ref.~\cite{Kramer:2001hh} \footnote{Note that both fits used CTEQ5L~\cite{Lai:1999wy} whereas we have used here CTEQ6L1, whose results are anyhow very close.}.
\item[Set II:] This set is based on LDMEs of $J/\psi,\psi(2S),\chi_c,\Upsilon(1S),\Upsilon(2S),\Upsilon(3S),\chi_{b}(1P),\chi_b(2P)$ presented in Ref.~\cite{Sharma:2012dy}. The contributions of $\chi_{b}(3P)$ have been ignored. Hence, we will set the LDMEs of $\chi_{b}(3P)$ to be zero. The fit was performed at LO in $\alpha_s$. The LHC, Tevatron and RHIC data were used to perform this combined fit.
\item[Set III:] This set is based on LDMEs extracted from NLO analyses, \ie\ the LDMEs of $J/\psi,\psi(2S),\chi_c$ from Ref.~\cite{Shao:2014yta} and those of $\Upsilon(nS),\chi_b(nP),n=1,2,3$ from Ref.~\cite{Han:2014kxa}. The CO LDMEs of charmonium are extracted from Tevatron data~\cite{Shao:2014yta}, while both Tevatron data and LHC data were used in Ref.~\cite{Han:2014kxa}.
\item[Set IV:] This set is based on  LDMEs for charmonium~\cite{Gong:2012ug} and bottomonium~\cite{Feng:2015wka} production based on other NLO analyses. They are determined by a combined fit to Tevatron and LHC data.
\end{enumerate}

\begin{table}[!hbtp] 
\begin{center}
\begingroup
\renewcommand{\arraystretch}{1.3}
\begin{tabular}{c|ccc}\footnotesize
          &$J/\psi+\Upsilon(1S)$ &$J/\psi+\Upsilon(2S)$ & $J/\psi+\Upsilon(3S)$ \\
\hline
Set I & $0.0604^{+0.357}_{-0.0496}$  & $0.0185^{+0.108}_{-0.0152}$ & $0.0158^{+0.0950}_{-0.0131}$ \\
Set II  &  $0.0948^{+0.591}_{-0.0826}$ & $0.0146^{+0.0868}_{-0.0222}$ & $6.28 \cdot 10^{-3}~^{+3.40\cdot 10^{-2}}_{-5.09\cdot 10^{-3}}$ \\
Set III & $0.0767^{+0.474}_{-0.0675}$ & $0.0205^{+0.116}_{-0.0179}$ & $1.14\cdot 10^{-2}~^{+6.34\cdot 10^{-2}}_{-1.01\cdot 10^{-2}}$ \\
Set IV & $0.0202^{+0.109}_{-0.0163}$ & $6.00\cdot 10^{-3}~^{+3.36\cdot 10^{-2}}_{-4.89\cdot 10^{-3}}$ & $2.51\cdot 10^{-3}~^{+1.34\cdot 10^{-2}}_{-2.03\cdot 10^{-3}}$ \\
\hline\hline
& $\psi(2S)+\Upsilon(1S)$ & $\psi(2S)+\Upsilon(2S)$ & $\psi(2S)+\Upsilon(3S)$\\
\hline
Set I &$1.85 \cdot 10^{-3}~^{+1.01\cdot 10^{-2}}_{-1.50\cdot 10^{-3}}$ & $5.83\cdot 10^{-4}~^{+3.15\cdot 10^{-3}}_{-4.72\cdot 10^{-4}}$ & $4.64\cdot 10^{-4}~^{+2.57\cdot 10^{-3}}_{-3.78\cdot 10^{-4}}$
\\
Set II & $4.30\cdot 10^{-3}~^{+2.62\cdot 10^{-2}}_{-3.73\cdot 10^{-3}}$ & $6.78\cdot 10^{-4}~^{+3.94\cdot 10^{-3}}_{-1.01\cdot 10^{-3}}$ & $3.09\cdot 10^{-4}~^{+1.64\cdot 10^{-3}}_{-2.49\cdot 10^{-4}}$\\
Set III & $3.19\cdot 10^{-3}~^{+1.98\cdot 10^{-2}}_{-2.84\cdot 10^{-3}}$ & $8.17\cdot 10^{-4}~^{+4.62\cdot 10^{-3}}_{-7.26\cdot 10^{-4}}$ & $4.57\cdot 10^{-4}~^{+2.54\cdot 10^{-3}}_{-4.11\cdot 10^{-4}}$ \\
Set IV & $9.03\cdot 10^{-4}~^{+4.78\cdot 10^{-3}}_{-7.30\cdot 10^{-4}}$ & $2.80\cdot 10^{-4}~^{+1.49\cdot 10^{-3}}_{-2.26\cdot 10^{-4}}$ & $1.42\cdot 10^{-4}~^{+6.81\cdot 10^{-4}}_{-1.13\cdot 10^{-4}}$ \\
\end{tabular}
\caption{$\sigma_{\rm SPS}(pp\to {\cal Q}_1+{\cal Q}_2) \times {\cal B}({\cal Q}_1\to\mu^+\mu^-){\cal B}({\cal Q}_2\to\mu^+\mu^-)$ in units of fb with $\sqrt{s}=115$ GeV, where ${\cal Q}_1=J/\psi,\psi(2S)$ and ${\cal Q}_2=\Upsilon(1S),\Upsilon(2S),\Upsilon(3S)$. We take four sets of LDMEs.}
\label{full}
\endgroup
\end{center}
\end{table}

\begin{table}[!hbtp] 
\begin{center}
\begingroup
\renewcommand{\arraystretch}{1.3}
\begin{tabular}{c|ccc}\footnotesize
          &$J/\psi+\Upsilon(1S)$ &$J/\psi+\Upsilon(2S)$ & $J/\psi+\Upsilon(3S)$ \\
\hline
$\{\ss,\so\}$ & $6.1\cdot 10^{-3}~^{+3.0\cdot 10^{-2}}_{-4.9\cdot 10^{-3}}$ & $1.8\cdot 10^{-3}~^{+8.6\cdot 10^{-3}}_{-1.4\cdot 10^{-3}}$ & $2.5\cdot 10^{-3}~^{+1.2\cdot 10^{-2}}_{-2.0\cdot 10^{-3}}$ \\
exclude feeddown  &  $0.024^{+0.15}_{-0.020}$ & $6.0\cdot 10^{-3}~^{+3.7\cdot 10^{-2}}_{-5.0\cdot 10^{-3}}$ & $0.011^{+0.065}_{-8.8\cdot 10^{-3}}$
 \\
include feeddown  &  $0.060^{+0.36}_{-0.050}$ & $0.019^{+0.11}_{-0.015}$ & $0.016^{+0.095}_{-0.013}$ \\
\hline\hline
& $\psi(2S)+\Upsilon(1S)$ & $\psi(2S)+\Upsilon(2S)$ & $\psi(2S)+\Upsilon(3S)$\\
\hline
$\{\ss,\so\}$ &$6.1\cdot 10^{-4}~^{+3.0\cdot 10^{-3}}_{-4.9\cdot 10^{-4}}$ & $1.8\cdot 10^{-4}~^{+9.0\cdot 10^{-4}}_{-1.5cdot 10^{-4}}$ & $2.4\cdot 10^{-4}~^{+1.2\cdot 10^{-3}}_{-1.9\cdot 10^{-4}}$\\
exclude feeddown  & $1.1\cdot 10^{-3}~^{+6.1\cdot 10^{-3}}_{-9.1\cdot 10^{-4}}$ & $3.0\cdot 10^{-4}~^{+1.6\cdot 10^{-3}}_{-2.4\cdot 10^{-4}}$ & $4.6\cdot 10^{-4}~^{+2.6\cdot 10^{-3}}_{-3.8\cdot 10^{-4}}$\\
include feeddown  & $1.9 \cdot 10^{-3}~^{+1.0\cdot 10^{-2}}_{-1.5\cdot 10^{-3}}$ & $5.8\cdot 10^{-4}~^{+3.2\cdot 10^{-3}}_{-4.7\cdot 10^{-4}}$ & $4.6\cdot 10^{-4}~^{+2.6\cdot 10^{-3}}_{-3.8\cdot 10^{-4}}$\\
\end{tabular}
\caption{$\sigma_{\rm SPS}(pp\to \Q_1+\Q_2) \times {\cal B}(\Q_1\to\mu^+\mu^-){\cal B}(\Q_2\to\mu^+\mu^-)$ in units of fb with $\sqrt{s}=115$ GeV, where $\Q_1=J/\psi,\psi(2S)$ and $\Q_2=\Upsilon(1S),\Upsilon(2S),\Upsilon(3S)$. We have used the Set I of the LDMEs. The uncertainty quoted comes only from the SDCs.}
\label{xsectionscbsetI}
\endgroup
\end{center}
\end{table}

In order to take into account the feeddown contributions, we have taken the necessary branching ratios from PDG~\cite{Agashe:2014kda}. For the unknown branching ratios, such as Br($\chi_{b}(3P)\rightarrow \Upsilon(nS)+\gamma$), we used the estimated values from Table I of Ref.~\cite{Han:2014kxa}.
The SPS cross sections of $\psi+\Upsilon$ production in proton-proton collisions at $\sqrt{s}=115$ GeV are presented in Table.~\ref{xsectionscbsps}. As clearly shown, the cross sections  significantly differ from one set of LDMEs to another. Before closing this appendix, we would like to stress several points.
\begin{itemize}
\item Because some CO LDMEs in Set II and Set IV are negative, the cross sections might be negative, which is of course unphysical. For example, the cross section for direct $J/\psi+\Upsilon(2S)$ production (which then excludes feeddowns) is negative for the  Set II and Set IV.
\item If one follows the arguments of Ref.~\cite{Ko:2010xy}, one is entitled to consider only the  $c\bar{c}(\so)+b\bar{b}[\so]$, $c\bar{c}[\ss]+b\bar{b}[\so]$ and $c\bar{c}[\so]+b\bar{b}[\ss]$ channels. This approximation is however based on the validity of the velocity-scaling rules of the LMDEs which may not be reliable. By using Set I of LDMEs, we have shown the comparison in Table.~\ref{xsectionscbsetI}. The row ${\ss,\so}$ only include $c\bar{c}(\so)+b\bar{b}[\so]$, $c\bar{c}[\ss]+b\bar{b}[\so]$ and $c\bar{c}[\so]+b\bar{b}[\ss]$ channels, while the remaining lines contain all CO and CS contributions (with or without feeddown contributions). The results clearly show that the $c\bar{c}(\so)+b\bar{b}[\so]$, $c\bar{c}[\ss]+b\bar{b}[\so]$ and $c\bar{c}[\so]+b\bar{b}[\ss]$ channels are not sufficient. Moreover, the feeddown contributions are also significant but for $\psi(2S)+\Upsilon(3S)$.
\item The CO LDMEs used in this section are mainly determined by data in the high transverse momentum region. It is important to point out that these LDMEs  yield to cross sections overestimating the data in the low transverse momentum region and, hence,  the total cross sections for the single quarkonium production (see e.g. Ref.~\cite{Feng:2015cba}). Hence, it is likely that any such NRQCD based estimation of $\psi+\Upsilon$ at low $P_T$ are too optimistic.
 However, as a conservative estimation, it is reasonable that we consider them as conservation upper limits of the SPS contributions (see Table.~\ref{xsectionscb}).
\item Finally, let us note that the relative importance of pure CO+CO contributions as compared to the mixed CO+CS depends much on the LDME sets. It essentially ranges from 30 to 70~\% irrespective of the charmonium-bottomonium pair which is considered. For the sake of completeness, let us add that the pure CS+CS from double feed-down from $\chi_c+\chi_b$ is on the order of a couple of per cent, but for Set IV where it can be up to 10\%.
\end{itemize}


\section*{Acknowledgements}
This work is 
supported in part by the France-China Particle Physics Laboratory (FCPPL) and by the French CNRS via the grants PICS-06149 Torino-IPNO,
FCPPL-Quarkonium4AFTER \& PEPS4AFTER2. H.-S. Shao is 
also supported by the ERC grant 291377 ``LHCtheory: Theoretical predictions and analyses of LHC physics: 
advancing the precision frontier".


\bibliographystyle{utphys}

\bibliography{doubleonia_AFTER_130815}

\end{document}